\documentclass[aps,prd,twocolumn,superscriptaddress,floatfix,longbibliography]{revtex4-2}
\usepackage[dvipsnames]{xcolor}
\usepackage{amssymb,amsmath,amsfonts,bm,slashed,soul,ragged2e,graphicx,epstopdf,hyperref,array}
\usepackage[utf8]{inputenc}
\usepackage{colortbl,color,epsfig,subfigure,caption}

\enlargethispage{\baselineskip}
\definecolor{steelblue}{RGB}{25,25,112}
\definecolor{dullblue}{rgb}{0,0.298,0.49}
\definecolor{darkred}{rgb}{0.545,0,0}
\definecolor{blue2}{cmyk}{1, 0.1, 0.1, 0}
\hypersetup{colorlinks,linkcolor={darkred},citecolor={blue},urlcolor={dullblue}}
\DeclareCaptionJustification{justified}{\justifying}
\everymath{\displaystyle}

\usepackage{comment}
\usepackage{braket}
\usepackage{amsmath}
\usepackage{soul}
\usepackage{makecell}
\usepackage{verbatim}
\usepackage{booktabs}
\usepackage{siunitx}

\usepackage{orcidlink}

\newcommand{\beq}{\begin{equation}}
\newcommand{\eeq}{\end{equation}}
\newcommand{\bea}{\begin{eqnarray}}
\newcommand{\eea}{\end{eqnarray}}

\newcommand{\gsim}{\lower.7ex\hbox{$\;\stackrel{\textstyle>}{\sim}\;$}}
\newcommand{\lsim}{\lower.7ex\hbox{$\;\stackrel{\textstyle<}{\sim}\;$}}

\newcommand{\be}{\begin{equation}}
\newcommand{\ee}{\end{equation}}
\newcommand{\ba}{\begin{eqnarray}}
\newcommand{\ea}{\end{eqnarray}}

\newcommand{\D}{\mathrm{d}}

\RequirePackage[normalem]{ulem}

\begin{document}

\title{
%Searching for Exotic Spin-Dependent Interactions with Adiabatic Spin Inversion
Probing Exotic Electron–Electron Interactions via Adiabatic Spin Inversion in MRFM}

\author{Chunlong Li}
\thanks{These authors contributed equally to this work.}
\affiliation{School of Physics, Hefei University of Technology, Hefei 230601, Anhui, China}
\author{Shien Yang}
\thanks{These authors contributed equally to this work.}
\affiliation{School of Physics, Hefei University of Technology, Hefei 230601, Anhui, China}
\author{Yiwei Huang}
\affiliation{School of Physics, Hefei University of Technology, Hefei 230601, Anhui, China}
\author{Yu Zhang\,\orcidlink{0000-0001-9415-8252}}
\email{dayu@hfut.edu.cn \,(Corresponding author)}
\affiliation{School of Physics, Hefei University of Technology, Hefei 230601, Anhui, China}
\author{Fei Xue\,\orcidlink{0000-0002-5681-229X}}
\email{xfei.xue@hfut.edu.cn \,(Corresponding author)}
\affiliation{School of Physics, Hefei University of Technology, Hefei 230601, Anhui, China}

\begin{abstract}

Light bosons beyond the Standard Model can mediate exotic spin-dependent interactions between electrons, motivating sensitive searches at short distances. We propose a magnetic resonance force microscopy (MRFM) scheme that exploits oscillating-cantilever-driven adiabatic reversals (OSCAR) of electron spins to convert these interactions into measurable mechanical signals. The ferromagnetic tip serves as both a dense polarized-spin source and a source of magnetic-field gradients, enabling force detection without magnetic shielding between the source and sensor. Within a linearized treatment of the coupled spin–mechanical dynamics, we derive resonance-frequency shifts associated with the \(V_2\) and \(V_3\) interactions and a second-harmonic force signal arising from the velocity-dependent \(V_8\) interaction. We further investigate interrupted-OSCAR differential detection and distance-dependent measurements to discriminate exotic signals from conventional magnetic backgrounds. Using representative parameters from ensemble and single-spin MRFM experiments, we estimate the projected sensitivity under specified noise and statistical assumptions. The results suggest that the proposed scheme could provide complementary sensitivity to short-range spin-dependent interactions, with the potential to improve existing constraints on \(V_2\) over selected submicrometer interaction ranges. Our study establishes adiabatic spin-inversion MRFM as a promising approach to probing exotic electron–electron forces.
\end{abstract}

\date{\today}

\maketitle

\section{Introduction}

The Standard Model (SM), despite its remarkable success, is known to be incomplete and cannot account for several established phenomena, such as neutrino masses and dark matter \cite{BERTONE2005279}. Light bosons beyond the SM provide generic mediators of new interactions and can generate a variety of exotic spin-dependent forces between Standard Model fermions \cite{PhysRevD.30.130,Bogdan_2006,RevModPhys.97.025005,Jiang_2025,liangyong2026exploring}. Depending on the mediator and coupling structure, these interactions can arise from axial--axial, vector--vector, or pseudoscalar--pseudoscalar couplings \cite{Langacker:2008yv,PhysRevLett.94.151802,PhysRevD.68.035012,annurev-nucl-102014-022120,PhysRevLett.38.1440} and are commonly parameterized by nonrelativistic potentials such as \(V_2\), \(V_3\), and \(V_8\). Various laboratory experiments have searched for macroscopic spin-dependent interactions between polarized electrons over a broad range of length scales. Existing constraints have been established using torsion-balance experiments \cite{PhysRevD.42.977,PhysRevLett.111.151802}, atomic magnetometers \cite{PhysRevLett.125.201802,PhysRevLett.121.261803}, solid-state spin systems \cite{Rong:2018yos}, trapped-ion platforms \cite{Kotler:2015ura}, atomic and molecular spectroscopy \cite{Ficek:2016qwp,Jiao:2019ikv}, and geoelectron-based experiments exploiting polarized electrons within the Earth \cite{doi:10.1126/science.1227460,PhysRevLett.112.091803}.

In searches for exotic spin-dependent interactions, the signal is typically modulated to distinguish it from environmental noise. However, the magnetic dipole-dipole interaction within the Standard Model is modulated in the same way, so that specially designed magnetic shielding is required to suppress this background. This severely constrains the achievable source-sensor separation, particularly in mechanically suspended systems \cite{PhysRevLett.134.111001,Yin:2025uzf,PhysRevLett.115.201801} and atomic comagnetometers \cite{PhysRevLett.125.201802,PhysRevLett.121.261803,PhysRevLett.129.051801,PhysRevLett.133.191801,doi:10.1126/sciadv.ade0353,doi:10.1126/sciadv.abi9535}, and limits sensitivity at sub-100~${\rm \mu m}$ length scales. Solid-state quantum spin sensors, such as NV centers in diamond, can circumvent bulky magnetic shielding through coherent spin control and quantum initialization and readout techniques, enabling searches at shorter distances \cite{Rong:2018yos,PhysRevLett.127.010501,Rong:2017wzk,PhysRevLett.131.071801,PhysRevLett.132.180801}. Nevertheless, their sensitivity is often limited by the relatively small number of controllable source or sensor spins. To overcome these limitations, we propose, for the first time, a magnetic resonance force microscopy (MRFM) scheme that exploits oscillating-cantilever-driven adiabatic reversals (OSCAR) of spins to convert exotic electron-electron interactions into measurable mechanical signals, with particular focus on the \(V_2\), \(V_3\), and \(V_8\) potentials arising from axial--axial, vector--vector, and pseudoscalar--pseudoscalar couplings.

MRFM is a hybrid spin-mechanical platform in which a ferromagnetic tip mounted on a mechanical resonator is magnetically coupled to a polarized spin sample under magnetic-resonance conditions \cite{PhysRevB.57.1108,Poggio_2010}. Conventionally, MRFM has been developed for ultrasensitive spin detection and nanoscale magnetic resonance imaging, including single-electron-spin detection and three-dimensional imaging of nuclear-spin ensembles \cite{mamin2007nuclear,degen2009nanoscale}. Magnetic cantilever-based force measurements have also been employed to search for exotic spin-dependent interactions \cite{PhysRevD.104.032008,PhysRevLett.124.161801}, whereas the present scheme exploits OSCAR spin inversions for force detection. The ferromagnetic tip provides an ultrahigh density of polarized electron spins, enhancing exotic spin-dependent interactions while generating the large magnetic-field gradient required for force detection. Meanwhile, the ordinary magnetic dipole interaction couples cantilever motion to the longitudinal spin polarization, enabling OSCAR spin inversions that map interaction-induced changes in spin polarization onto the mechanical response of the cantilever \cite{PhysRevLett.91.207604,PhysRevLett.87.277602,rugar2004single,PhysRevA.66.023405,SUTER2002210,peddibhotla2013harnessing,PhysRevB.84.205328}. These characteristics overcome the limited polarized-spin density of existing solid-state sensors while eliminating the need for magnetic shielding between the source and sensor, and provide the basis for long-interrogation-time measurements and force-based signal discrimination. Together, they make MRFM a promising platform for high-sensitivity searches for spin-dependent interactions at submicrometer length scales.

\section{Magnetic Resonance Force System}

\begin{figure*}
    \centering
    \includegraphics[width=0.95\linewidth]{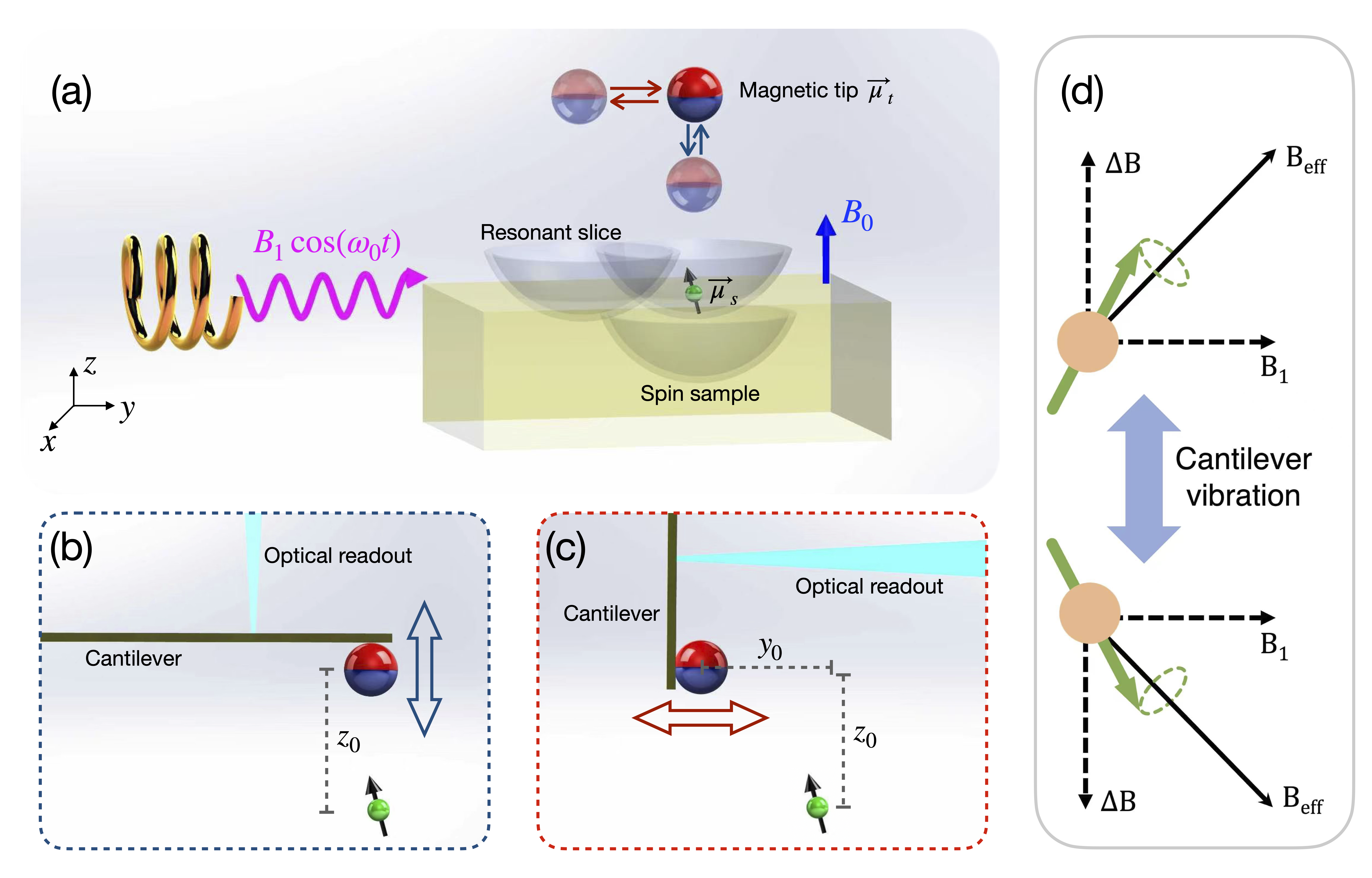}
    \caption{Schematic of the MRFM configurations considered in this work. (a) Generic setup consisting of a ferromagnetic tip, a spin sample, a static bias field \(\vec B_0\), and a microwave field \(B_1\cos(\omega_0 t)\). The inhomogeneous tip field defines a resonant slice in the sample. (b) Longitudinal geometry, where the tip oscillates along the tip--sample separation direction about an equilibrium separation \(z_0\), and the resonant slice sweeps through a finite sample depth. (c) Transverse geometry, where the tip motion is transverse to the tip--sample separation, with equilibrium offsets \(y_0\) and \(z_0\). (d) Adiabatic spin inversion in the rotating frame: cantilever motion modulates the detuning \(\Delta B\), causing \(\vec B_{\rm eff}\) to reverse its longitudinal component and the sample spin to adiabatically follow.}
    \label{fig:setup}
\end{figure*}

As illustrated in Fig.~\ref{fig:setup}(a), a typical MRFM system consists of a ferromagnetic tip attached to a mechanical cantilever and positioned in close proximity to a spin sample, together with a static bias field \(\vec B_0\) and a microwave field \(B_1\cos(\omega_0 t)\) for spin manipulation. The tip generates a strongly inhomogeneous magnetic field \(\vec B_t\). Because the longitudinal component \(B_t^z(\vec r)\) varies strongly in space, only spins satisfying the local resonance condition $\omega_0=\gamma\left[B_0+B_t^z(\vec r)\right]$ within the resonance linewidth are efficiently driven by the microwave field. These spins define a thin spatial region known as the resonant slice. As the cantilever oscillates, the tip field periodically modulates the position of the resonant slice and drives adiabatic reversals of the longitudinal spin polarization. The resulting magnetic-resonance force acts back on the cantilever as an effective spring-constant correction, shifting the resonance frequency from \(\omega_c\) to \(\omega_c^*\). The cantilever dynamics can then be written as
\begin{align}
    \ddot{q}+\frac{\omega_c^*}{Q^*}\dot{q}+{\omega_c^*}^2 q=\frac{F_i}{m_{\rm eff}},
\end{align}
where \(q\) denotes the displacement of the tip micromagnet along the relevant cantilever mode, \(Q^*\) is the corresponding quality factor, and \(m_{\rm eff}\) is the effective mass of the cantilever--tip oscillator. The additional force \(F_i\) represents the contribution from a possible exotic spin-dependent interaction between the tip and the sample spins.

We consider two representative MRFM geometries that have been realized in classic experiments. Fig.~\ref{fig:setup}(b) shows an longitudinal geometry, in which the tip motion is mainly along the tip--sample separation direction and the resonant slice sweeps through a finite sample depth, allowing multiple spins to contribute to the signal \cite{PhysRevLett.87.277602}. Fig~\ref{fig:setup}(c) shows a transverse geometry, in which the cantilever motion is predominantly transverse to the tip--sample separation, representative of the single-electron MRFM experiment of Rugar \textit{et al.} \cite{rugar2004single}.

\section{Signal of exotic interactions}

Possible new bosons beyond the Standard Model can mediate spin-dependent interactions between electrons. In the nonrelativistic limit, these interactions can be expressed in terms of a set of effective potentials characterized by their spin, spatial, and velocity dependence \cite{Bogdan_2006,Jiang_2025,RevModPhys.97.025005}. We first consider the axial--axial interactions mediated by a massive spin-1 boson $Z^\prime$. For a sample electron spin $\vec{\sigma}_s$ interacting with a polarized electron spin $\vec{\sigma}_t$ in the ferromagnetic tip, the potentials relevant to the present MRFM scheme are
\begin{align}
    V_2^{AA}&=-g_A^e g_A^e \frac{\hbar c}{4 \pi} \vec{\sigma}_t \cdot \vec{\sigma}_s \frac{1}{r} e^{-r / \lambda}, \nonumber \\
    V_3^{AA}&=-g_A^eg_A^e\frac{\hbar c}{4\pi}\Bigg[(\vec\sigma_t\cdot\vec\sigma_s)A(r)-(\vec\sigma_t\cdot\hat r)(\vec\sigma_s\cdot\hat r)C(r)\Bigg], \nonumber \\
    V_8^{AA}&=-g_A^e g_A^e \frac{\hbar}{8 \pi c}\left(\vec{\sigma}_t \cdot \vec{v}\right)\left(\vec{\sigma}_s \cdot \vec{v}\right) \frac{1}{r} e^{-r / \lambda},
\end{align}
where
\begin{align}
    A(r)&=\left(\frac{\lambda^2}{r^3}+\frac{\lambda}{r^2}\right)e^{-r/\lambda}, \\
    C(r)&=\left(\frac{3\lambda^2}{r^3}+\frac{3\lambda}{r^2}+\frac1{r}\right)e^{-r/\lambda},
\end{align}
and \(\lambda=\hbar/(m_{Z^\prime}c)\) is the Compton wavelength of the exchanged mediator, which sets the interaction range. Here, \(g_A^e\) is the axial-vector coupling to electrons, \(r\) is the separation between the source and sample electrons, and \(\vec v\) is their relative velocity. The contact term in \(V_3^{AA}\) is omitted because the source and sensor spins are spatially separated in the present MRFM geometry \cite{RevModPhys.97.025005}.

The same spin and spatial structures also arise for vector--vector and pseudoscalar--pseudoscalar couplings. For an electron--electron system, the corresponding \(V_2^{VV}\), \(V_3^{VV}\), and \(V_3^{pp}\) interactions can therefore be obtained from \(V_2^{AA}\) and \(V_3^{AA}\) by
\begin{align}
    V_{2,3}^{VV}:\qquad &g_A^eg_A^e\longrightarrow-g_V^eg_V^e\,\eta_e(\lambda'), \nonumber \\
    V_3^{pp}:\qquad &g_A^eg_A^e\longrightarrow g_p^eg_p^e\,\eta_e(\lambda'),
    \label{rescale}
\end{align}
with
\begin{align}
    \eta_e(\lambda')=\frac{1}{4}\left(\frac{\lambda_e}{\lambda'}\right)^2, \lambda_e=\frac{\hbar}{m_ec}.
\end{align}
Here \(\lambda'\) denotes the Compton wavelength of the corresponding exchanged mediator, and is therefore mediator dependent.

We next evaluate the mechanical signals generated by these interactions in the MRFM configurations introduced above. Treating the ferromagnetic tip as a continuous distribution of polarized electron spins with number density \(n_t\), and retaining the leading contribution in the cantilever displacement, the exotic forces along the mechanical mode can be written as
\begin{align}
    F_{2,3}&=g_A^eg_A^e\frac{3\hbar c \mu_0}{8\pi^2 \mu_B} \frac{\mu_t\mu_s n_t}{B_1} I_{2,3}(\vec{r}_0,\lambda) q, \\
    F_8&=g_A^eg_A^e\frac{3\hbar\mu_0}{8\pi^2 c \mu_B} \frac{\mu_t\mu_s n_t}{B_1} I_8(\vec{r}_0,\lambda) \left( q\ddot{q}+\frac12\dot{q}^2 \right),
\end{align}
where the \(V_2\) and \(V_3\) contributions are grouped as \(F_{2,3}\) owing to their identical functional forms. Both \(F_{2,3}\) and \(F_8\) arise from OSCAR-driven adiabatic spin inversion (Fig.~\ref{fig:setup}(d)), under which the longitudinal spin polarization follows the cantilever displacement, \(\mu_s^z\propto q\), in the linear-response regime, as derived in Appendix~A. Here, \(\mu_t\) and \(\mu_s\) denote the magnetic moments of the tip and sample spins, respectively, \(B_1\) is the microwave-field amplitude, and \(\vec r_0\) specifies the equilibrium tip--spin configuration. The factors \(I_i(\vec r_0,\lambda)\) encode two distinct ingredients: the experimental geometry, specified by the equilibrium tip--spin configuration (Fig.~\ref{fig:setup}(b) and (c)), and the spatial kernel determined by the corresponding interaction potential \(V_i\) and its range \(\lambda\). Their explicit forms for the longitude and transverse geometries are given in Appendix~B. Since \(F_2\) and \(F_3\) are linear in the cantilever displacement, they act as corrections to the effective spring constant and therefore shift the mechanical resonance frequency,
\begin{align}
    \Delta\omega_c^*=-g_A^eg_A^e\frac{3\hbar c \mu_0}{16\pi^2 \mu_B} \frac{\mu_t\mu_s n_t}{m_{\rm eff}\omega_c^* B_1} I_{2,3}(\vec{r}_0,\lambda).
\end{align}
The \(V_2\) and \(V_3\) interactions can thus be probed through precision measurements of the cantilever resonance frequency, with their spatial dependences contained in \(I_2\) and \(I_3\). By contrast, the velocity-dependent \(V_8\) interaction does not directly induce a resonance-frequency shift. For nearly harmonic motion \(q(t)=q_a\cos(\omega_c t)\), the term \(q\ddot q+\dot q^2/2\) generates a component at \(2\omega_c\). Consequently, \(V_8\) produces a second-harmonic force component \(F_{8,2\omega_c}\), providing a signal channel distinct from the frequency shifts generated by \(V_2\) and \(V_3\). This response vanishes in the transverse geometry considered here, where the cantilever velocity is perpendicular to the spin polarization.

\section{Background subtraction}
\label{sec:bs}

In experiments employing ferromagnetic materials for searches of spin-dependent new physics, controlling backgrounds arising from the magnetic source is a critical challenge \cite{Gramolin:2020ict}. These backgrounds fall into two categories. The first consists of spin-independent forces, such as Casimir forces, electrostatic forces, and forces arising from the magnetization induced by the stray field of the tip. The second is the spin-dependent magnetic resonance force originating from the conventional magnetic dipole interaction.

\begin{figure}
    \centering
    \includegraphics[width=0.95\linewidth]{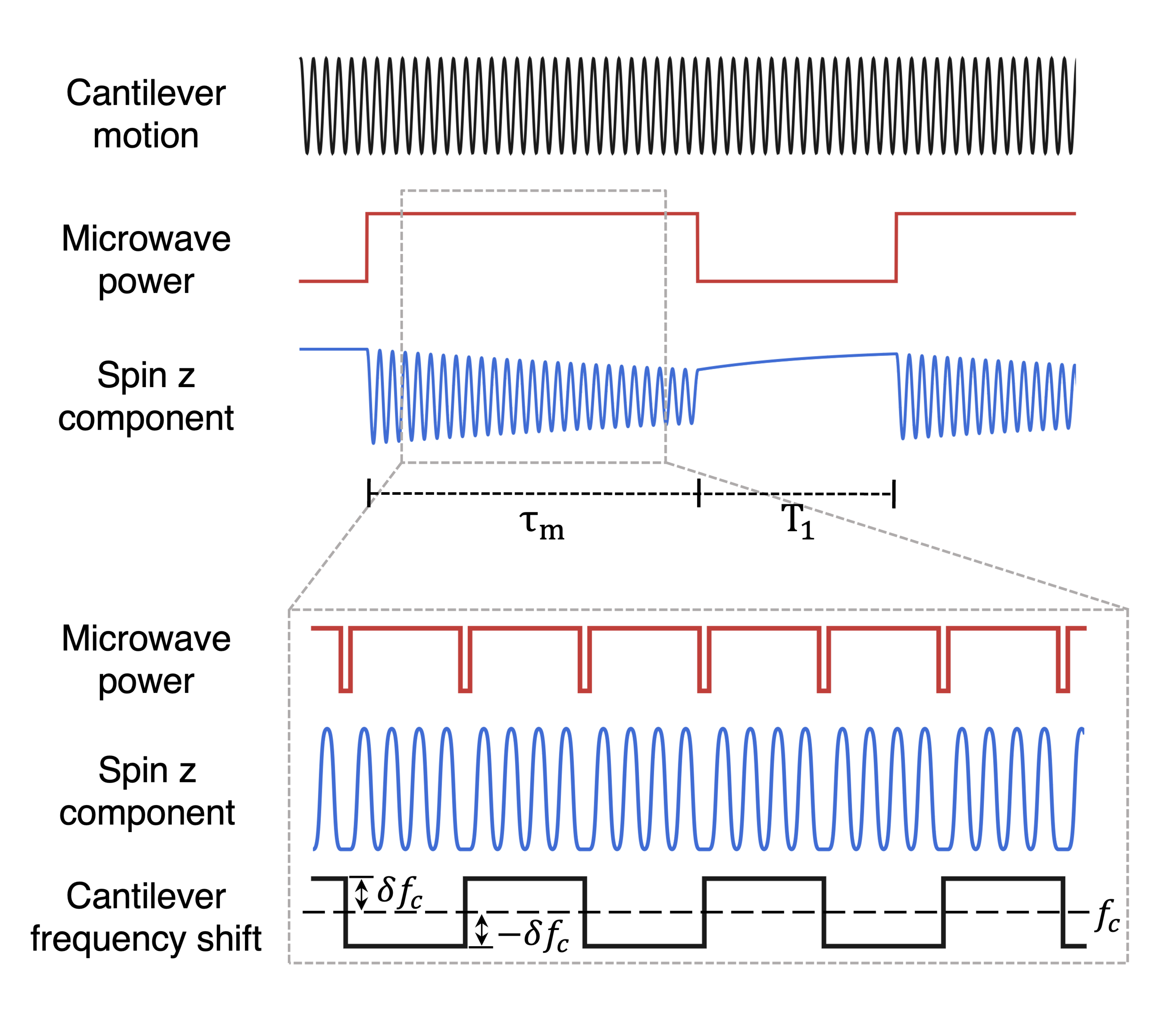}
    \caption{Measurement protocol for background suppression and sensitivity estimation. The microwave field is applied for a synchronization interval \(\tau_m\), during which the sample-spin \(z\) component follows the cantilever motion and produces a coherent magnetic-resonance force, followed by a repolarization interval \(T_1\). The lower panel enlarges the iOSCAR sequence within \(\tau_m\): periodic interruptions of the microwave excitation reverse the phase of the spin-dependent force, modulating the cantilever frequency shift between \(+\delta f_c\) and \(-\delta f_c\), while spin-independent backgrounds remain unmodulated.}
    \label{fig:protocal}
\end{figure}

To distinguish exotic-interaction signals from these backgrounds, we consider the measurement protocol illustrated in Fig.~\ref{fig:protocal}. As shown in the upper panel, each measurement cycle begins with a synchronization interval \(\tau_m\), during which the sample spins remain synchronized with the cantilever motion and generate a coherent magnetic-resonance force. Once synchronization is lost, the microwave excitation is switched off, allowing the spins to repolarize under the static magnetic field \(B_0\) over a recovery interval characterized by the spin-lattice relaxation time \(T_1\), before the next measurement cycle begins.

Building on the OSCAR adiabatic spin-inversion scheme introduced above, spin-independent backgrounds can be suppressed using the interrupted OSCAR (iOSCAR) protocol, as illustrated in Fig.~\ref{fig:protocal} \cite{PhysRevLett.91.207604}. By periodically interrupting the microwave excitation, iOSCAR switches between opposite adiabatic spin-inversion phases while leaving spin-independent forces essentially unchanged. Denoting the two adiabatic branches by \(\eta=\pm1\), the longitudinal spin polarization satisfies \(\mu_s^z(\eta)\propto\eta q\), leading to
\begin{align}
    \Delta\omega_c^{*(\eta)}&=\eta\,\Delta\omega_c^*, \nonumber\\
    F_{8,2\omega_c}^{(\eta)}&=\eta\,F_{8,2\omega_c}.
    \label{ioscar_signals}
\end{align}
Both the resonance-frequency shift and the second-harmonic force component reverse sign upon switching between the two adiabatic branches. Thus, phase-sensitive differential detection can suppress spin-independent backgrounds in both channels, including nonlinear forces that may also generate a response at \(2\omega_c\).

After suppressing spin-independent backgrounds, the remaining signals contain contributions from both the conventional magnetic interaction and exotic spin-dependent interactions. These can be further distinguished through their different distance dependences by measurements at multiple characteristic source--sensor separations $d$ \cite{Rong:2018yos,Jiao:2019ikv}. Here, $d$ is a generic separation parameter whose relation to the equilibrium coordinates, such as $z_0$ or $(y_0,z_0)$, depends on the MRFM geometry. Fitting the distance-dependent frequency shifts or second-harmonic responses separates the conventional magnetic and exotic contributions, enabling constraints on the corresponding coupling strengths (Appendix C).

\section{Projected sensitivity}

Following the measurement protocol described in Sec.~\ref{sec:bs}, the effective integration time over a total measurement duration \(T_{\rm tot}\) is estimated as \(T_{\rm int}\approx[\tau_m/(\tau_m+T_1)]T_{\rm tot}\), where the factor \(\tau_m/(\tau_m+T_1)\) accounts for the fraction of time spent on coherent signal acquisition. The resulting sensitivity depends on how the cantilever force noise enters the measurement of the respective signal observables. For the \(V_{2,3}\) search, the signal appears as a resonance-frequency shift of the cantilever, and the sensitivity is therefore limited by the resonance-frequency uncertainty \(\sigma_\omega\), which is evaluated using the phase-sensitive frequency-estimation method developed in previous MRFM experiments \cite{PhysRevLett.91.207604,PhysRevLett.87.277602,rugar2004single}. For the \(V_8\) search, the signal is identified as a spectral peak at \(2\omega_c\), and the sensitivity is instead determined by the minimum detectable force \(\sigma_F\) obtained from the force-noise power spectrum.

To evaluate \(\sigma_\omega\) and \(\sigma_F\), we first analyze the force-noise power spectral density \(S_F\) acting on the cantilever. The total noise consists of two contributions: optical readout noise and intrinsic force noise. In an optimized MRFM setup, the optical readout noise is designed to remain below the intrinsic noise floor at the cantilever resonance frequency \(\omega_c\). For the \(V_8\) search, we additionally assume that the optical readout remains thermal-noise limited at \(2\omega_c\). Since the mechanical susceptibility is substantially smaller at this off-resonant frequency, achieving this condition requires sufficiently low displacement-imprecision noise. Under these assumptions, the intrinsic force noise includes the thermal force noise of the cantilever and the force noise associated with stochastic spin fluctuations in the sample, with the thermal contribution dominating for the ensemble configurations considered here (Appendix D). We therefore approximate the total force-noise PSD as \(S_F(\omega)\simeq S_F^{\rm th}=4k_BTm_{\rm eff}\omega_c/Q\), which is white over the frequency range of interest. The resulting sensitivity projections represent idealized thermal-noise-limited estimates. To provide an intuitive estimate of the achievable coupling reach, we consider the \(V_2\) interaction using the detection configuration illustrated in Fig.~\ref{fig:setup}(b). In the long-range limit \(\lambda\gg z_0\), the sensitivity approaches a constant value, yielding the following analytical estimate:
\begin{align}
    g_A^eg_A^e & \approx 10^{-17} \left(\frac{B_1}{1 {\rm mT}}\right) \left(\frac{m_{\rm eff}}{10^{-7}{\rm g}}\right)^{\frac12} \left(\frac{T}{4{\rm K}}\right)^{\frac12} \nonumber \\
    & \times \left(\frac{Q}{10^6}\right)^{-\frac12} \left(\frac{\mu_t}{10^{-12}{\rm J/T}}\right)^{-2} \left(\frac{N_s}{10^2}\right)^{-1}\left(\frac{z_0}{1{\rm \mu m}}\right)^6 \nonumber \\
    & \times \left( \frac{z_a}{10{\rm nm}}\right)^{-1} \left(\frac{f_c}{5{\rm kHz}}\right)^{\frac12} \left(\frac{T_{\rm int}}{10^3{\rm s}}\right)^{-\frac12}.
    \label{V2ana}
\end{align}
Here \(N_s\) denotes the effective number of polarized electron spins coupled to the magnetic tip through magnetic resonance, with \(N_s=1\) for single-spin detection and \(N_s\sim5\times10^2\) for spin-ensemble detection \cite{PhysRevLett.87.277602}.

\begin{figure}
    \centering
    \includegraphics[width=0.95\linewidth]{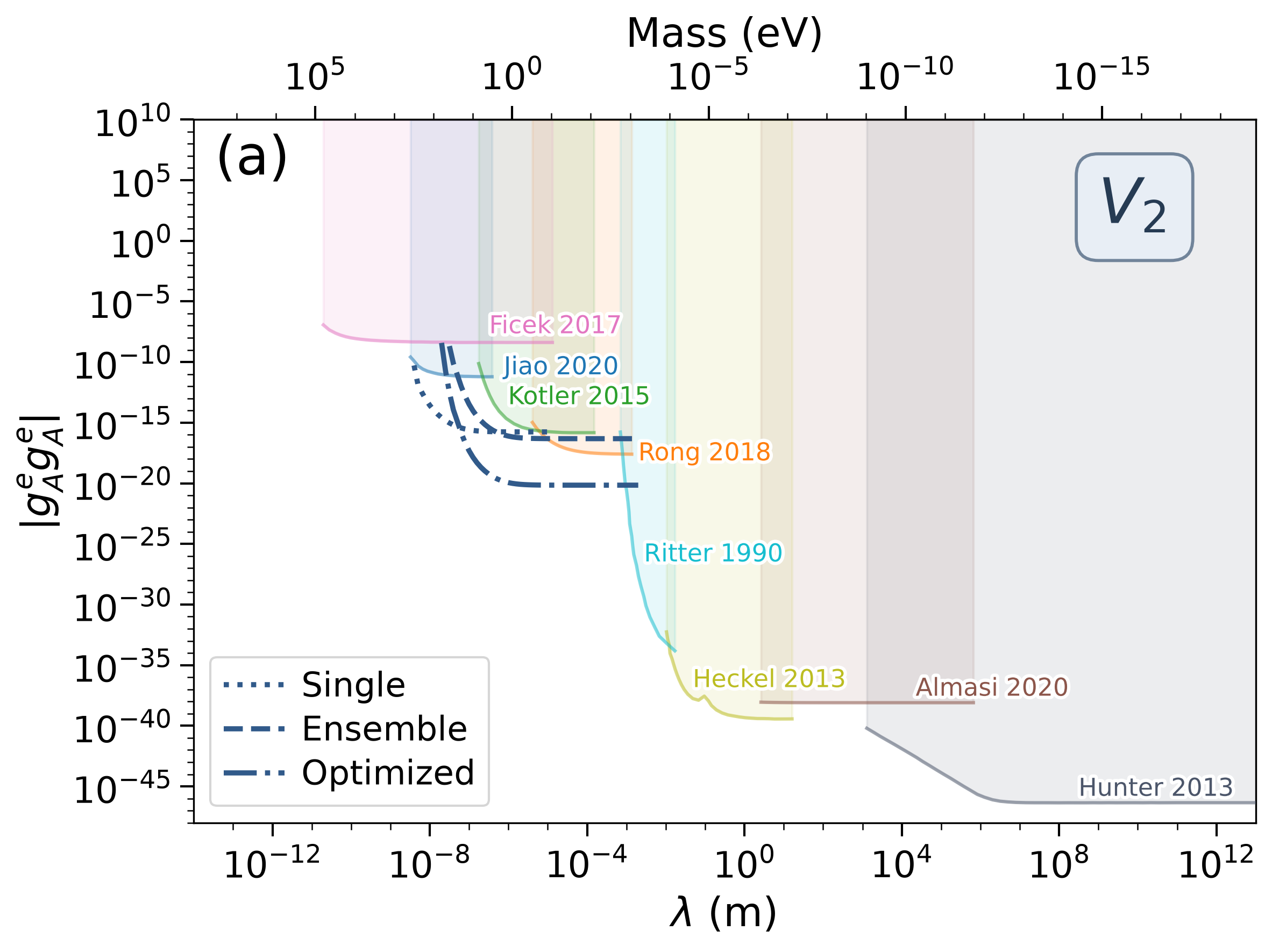}
    \includegraphics[width=0.95\linewidth]{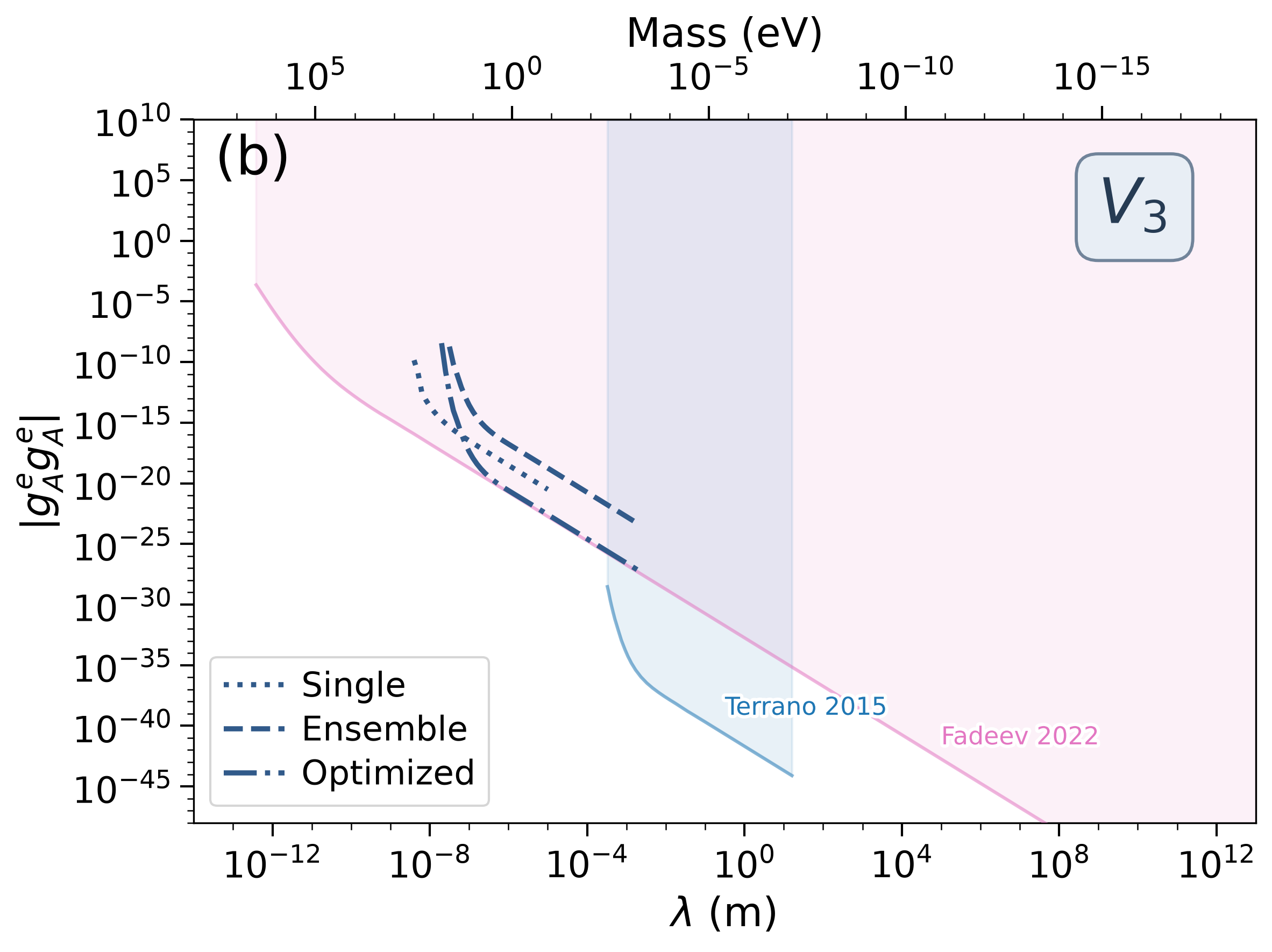}
    \includegraphics[width=0.95\linewidth]{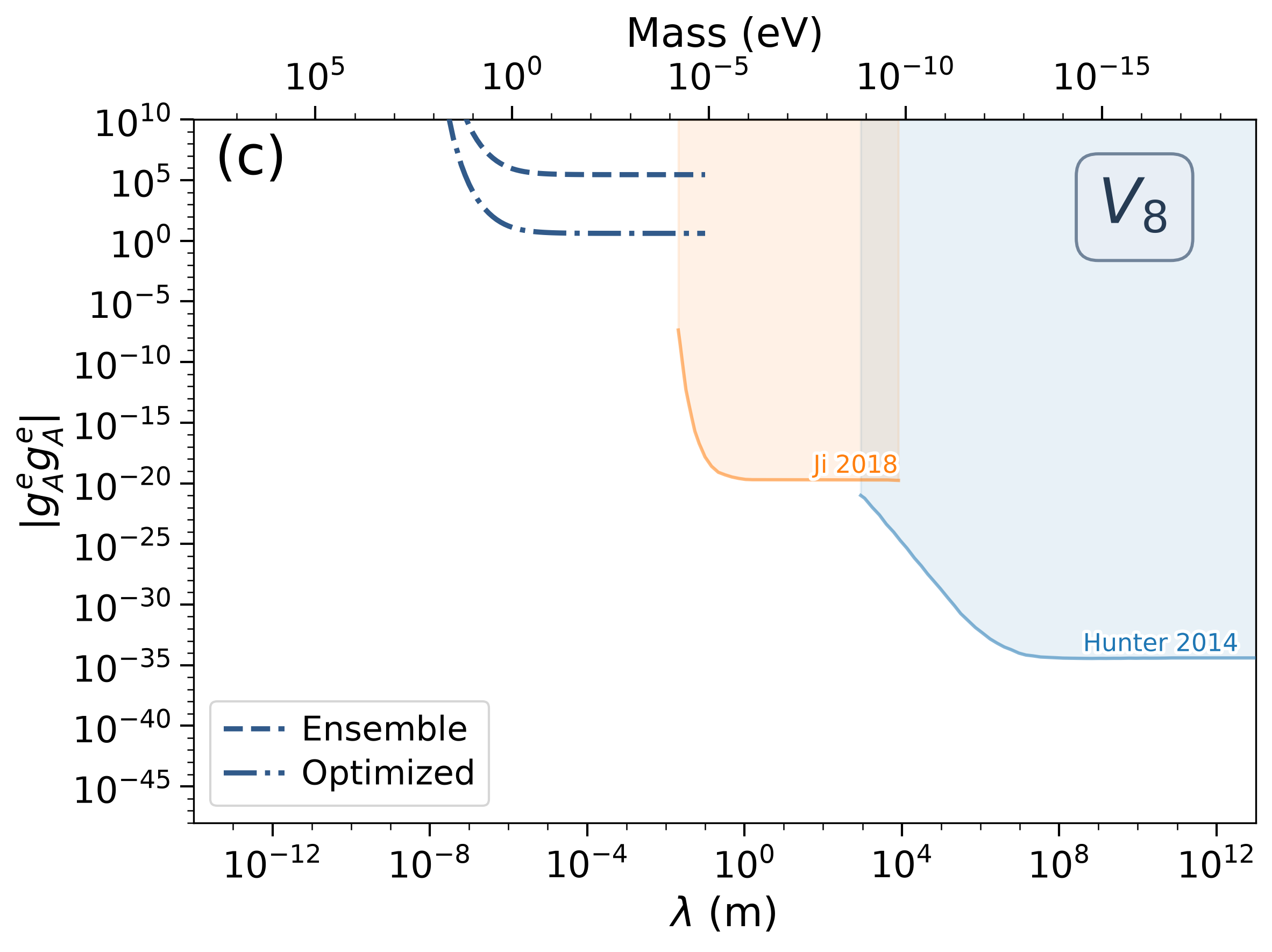}
    \caption{Projected MRFM constraints on \(g_A^eg_A^e\) from (a) \(V_2^{AA}\), (b) \(V_3^{AA}\), and (c) \(V_8^{AA}\). 
    Dotted, dashed, and dash-dotted curves denote the single-spin, ensemble, and optimized configurations, respectively.
    The ensemble case corresponds to approximately \(500\) polarized electron spins, while the “Optimized” curve assumes \(T=30~\mathrm{mK}\) and a cantilever oscillation amplitude \(z_a=100~\mathrm{nm}\). Existing experimental bounds are shown for comparison \cite{Ficek:2016qwp,Jiao:2019ikv,Kotler:2015ura,Rong:2018yos,PhysRevD.42.977,PhysRevLett.111.151802,PhysRevLett.125.201802,doi:10.1126/science.1227460,PhysRevLett.115.201801,PhysRevLett.121.261803,PhysRevLett.112.091803}.}
    \label{fig:prospect}
\end{figure}

To estimate the projected sensitivity under experimentally realistic conditions, we adopt representative parameters from demonstrated MRFM experiments. Specifically, we consider two configurations based on existing spin-ensemble and single-spin implementations, corresponding to the longitudinal geometry in Fig.~\ref{fig:setup}(b) \cite{PhysRevLett.87.277602} and the transverse geometry in Fig.~\ref{fig:setup}(c) \cite{rugar2004single}, respectively. The parameters used in the sensitivity analysis are summarized in Table~\ref{tab:parameters}, with their derivations and justifications provided in Appendix E. The resulting projected constraints on the axial-vector electron coupling \(g_A^eg_A^e\) are shown in Fig.~\ref{fig:prospect} for the \(V_2\), \(V_3\), and \(V_8\) interactions, together with existing experimental bounds \cite{RevModPhys.97.025005}.

\begin{table}[htbp]
\centering
\renewcommand{\arraystretch}{1.3}
\begin{tabular}{lll}
\hline
\hline
 & Ensemble & Single \\ \hline
Resonant frequency \(f_c\) & \(21.4\ \text{kHz}\) & \(5.5\ \text{kHz}\) \\
Quality factor \(Q\) & \(10^5\) & \(5\times 10^4\) \\
Effective mass \(m_{\mathrm{eff}}\) & \(3.0 \times 10^{-10}\ \text{g}\) & \(3.4 \times 10^{-14}\ \text{g}\) \\
Background field \(B_0\) & \(0.214\ \text{T}\) & \(0.106\ \text{T}\) \\
Microwave field \(B_1\) & \(3\ \text{G}\) & \(3\ \text{G}\) \\
Oscillation amplitude \(z_a\) & \(10\ \text{nm}\) & \(16\ \text{nm}\) \\
Temperature \(T\) & \(3\ \text{K}\) & \(1.6\ \text{K}\) \\
Tip radius \(r_t\) & \(1\ \mu\text{m}\) & \(0.1\ \mu\text{m}\) \\
Tip moment \(\mu_t\) & \(1.5\times 10^{-12} \text{J/T}\) & \(2.9\times 10^{-15} \text{J/T}\) \\
Tip–sample separation \(z_0\) & \(1500\ \text{nm}\) & \(125\ \text{nm}\) \\
Spin density \(n_e\) & \(2\times 10^{18}\ \text{cm}^{-3}\) & \(10^{14}\ \text{cm}^{-3}\) \\
Synchronization time \(\tau_m\) & \(275\ \text{ms}\) & \(760\ \text{ms}\) \\
Relaxation time \(T_1\) & \(13\ \text{s}\) & \(2\ \text{s}\) \\
Measurement time \(T_{\rm tot}\) & \(10^{4}\ \text{s}\) & \(10^{4}\ \text{s}\) \\
\hline
\hline
\end{tabular}
\caption{Experimental parameters used in our calculations. For the transverse configuration, \(y_0=0.39z_0\) is chosen to maximize the magnitude of the transverse magnetic-field gradient.}
\label{tab:parameters}
\end{table}

For \(V_2\), both the single-spin and spin-ensemble configurations are projected to improve existing constraints by several orders of magnitude over selected submicrometer interaction ranges. This sensitivity arises from the high polarized-spin density of the ferromagnetic tip, the short source--sensor separation, and the force-gradient-based readout of MRFM. Despite the much smaller tip--sample separation in the single-spin configuration, its projected sensitivity remains comparable to that of the ensemble setup, owing to the collective contribution of resonant spins in the latter. In the longitudinal ensemble configuration of Fig.~\ref{fig:setup}(b), the cantilever oscillation sweeps a resonant volume \(V\simeq2z_aA\), where \(z_a\) is the oscillation amplitude and \(A\) is the cross-sectional area of the resonant slice. The effective number of polarized spins is \(N_s=n_eVP\), where \(n_e\) is the sample spin density and \(P=\tanh(\mu_BB_0/k_BT)\) is the equilibrium spin polarization \cite{PhysRevLett.87.277602}. Further improvement by reducing the separation is limited by the finite tip size, since shrinking the tip also decreases its magnetic moment, \(\mu_t\propto r_t^3\), and the number of resonant spins \(N_s\). Alternatively, lowering the operating temperature and increasing the cantilever oscillation amplitude offer promising routes to enhanced sensitivity. Although the polarization of individual spins eventually saturates, a larger oscillation amplitude can increase the collective spin moment by enlarging the resonant volume, provided efficient adiabatic spin inversions are maintained (Appendix~E). The curves labeled “Optimized” illustrate a prospective ensemble configuration with \(T=30~\mathrm{mK}\) and \(z_a=100~\mathrm{nm}\), while all other independent experimental parameters are assumed to remain unchanged.

For \(V_3\), although the projected sensitivity does not surpass the strongest existing bounds from atomic spectroscopy \cite{Ficek:2016qwp,PhysRevA.105.022812}, MRFM provides an independent, force-based probe through adiabatic spin inversion. Rather than relying solely on comparisons between experimental measurements and Standard Model QED predictions, its tunable source--sensor separation enables direct measurements of the interaction's spatial dependence, allowing the exotic signal to be distinguished from conventional magnetic backgrounds and reducing associated systematic uncertainties. Recent helium spectroscopy has also provided complementary constraints on the combined axial-vector interaction \(V_2^{AA}+V_3^{AA}\) \cite{Cong:2026kuv}.

For the velocity-dependent \(V_8\) interaction, the sensitivity is intrinsically weaker because of the velocity suppression of the potential; nevertheless, the proposed scheme extends direct sensitivity to \(V_8\) into the submicrometer regime, well below the characteristic length scales of existing searches.

\begin{figure}
    \centering
    \includegraphics[width=0.95\linewidth]{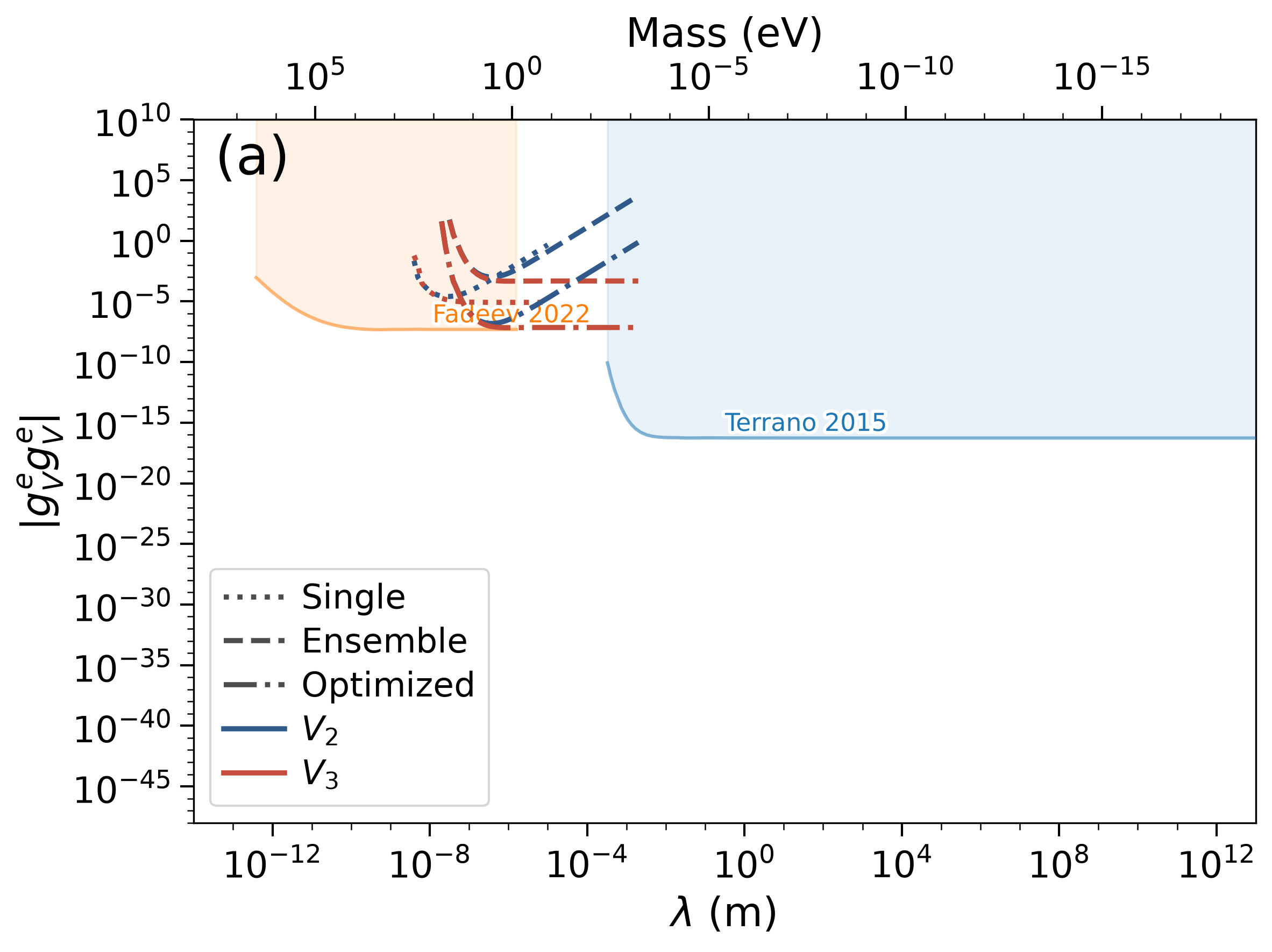}
    \includegraphics[width=0.95\linewidth]{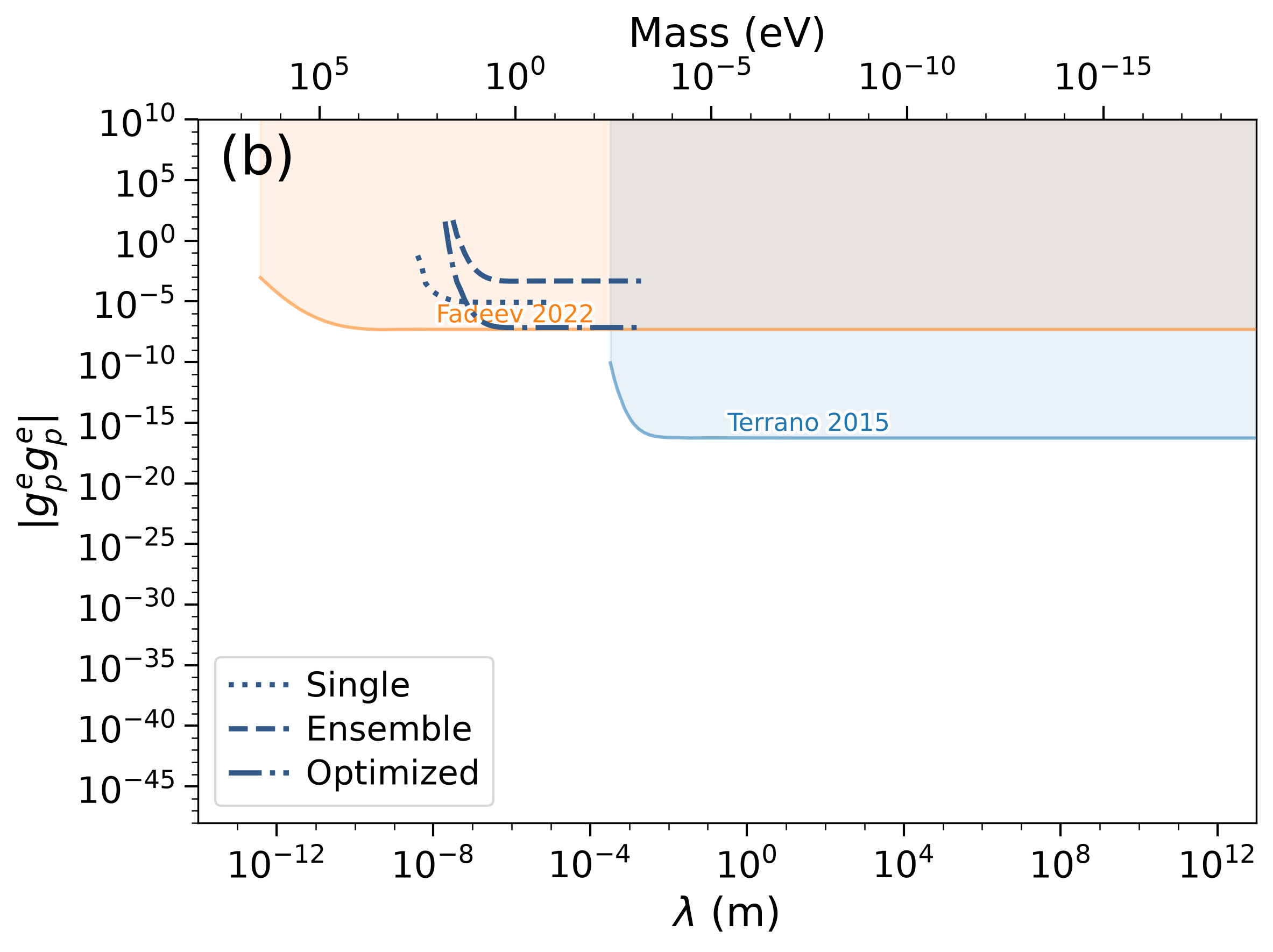}
    \caption{Projected MRFM constraints on (a) \(g_V^eg_V^e\) from \(V_2^{VV}\) and \(V_3^{VV}\), and (b) \(g_p^eg_p^e\) from \(V_3^{pp}\), obtained from the axial--axial results through Eq.~\eqref{rescale}. Dotted, dashed, and dash-dotted curves denote the single-spin, ensemble, and optimized configurations, respectively; existing bounds are shown for comparison \cite{PhysRevA.105.022812,PhysRevLett.115.201801}.}
    \label{fig:prospect_others}
\end{figure}

The axial--axial sensitivities can be translated into constraints on vector--vector and pseudoscalar--pseudoscalar couplings using Eq.~\eqref{rescale}, as shown in Fig.~\ref{fig:prospect_others}. For the vector--vector interaction, the baseline single-spin and ensemble configurations provide complementary sensitivity in the intermediate region between the two bounds, while the optimized ensemble scenario further improves the projected reach. For the pseudoscalar--pseudoscalar interaction, the baseline sensitivities remain weaker than existing constraints, whereas the optimized ensemble configuration can approach the current bounds over a limited interaction-range window.

\section{Discussions and conclusions}

Compared with existing approaches, the proposed MRFM scheme offers three main advantages for probing short-range spin-dependent interactions. First, existing searches in the micrometer and submicrometer regime include molecular-ruler measurements \cite{Jiao:2019ikv}, trapped-ion experiments \cite{Kotler:2015ura}, and NV-center-based solid-state spin sensors \cite{Rong:2018yos}. MRFM differs in that its ferromagnetic tip provides an ultrahigh-density polarized-spin source containing \(10^8\!-\!10^{11}\) polarized electron spins at submicrometer distances. Second, MRFM detects the longitudinal spin polarization through adiabatic spin inversion, so the effective interrogation time is governed by the synchronization time \(\tau_m\sim T_{1\rho}\) \cite{PhysRevB.57.1108}; experimentally demonstrated synchronization times can reach tens of milliseconds to nearly one second \cite{PhysRevLett.91.207604,PhysRevLett.87.277602,rugar2004single,RevModPhys.67.249}, exceeding the typical \(T_2\)-limited interrogation times of pulsed NV sensing \cite{RevModPhys.92.015004}. This long interrogation time has also motivated recent MRFM proposals for axion dark-matter searches \cite{kashi2025searching}. Third, force-based readout measures the spatial gradient of the exotic interaction rather than the interaction energy itself, leading to a steeper distance dependence for Yukawa-type forces and thereby improving discrimination against conventional magnetic backgrounds through distance-scaling measurements.

In summary, we have introduced a new application of cantilever-driven adiabatic spin reversals in MRFM to search for exotic electron--electron spin-dependent interactions mediated by light bosons beyond the Standard Model. By analyzing the coupled spin--mechanical dynamics, we identify two characteristic experimental signatures: resonance-frequency shifts induced by the \(V_2\) and \(V_3\) interactions and a second-harmonic force signal generated by the velocity-dependent \(V_8\) interaction. The same framework applies to axial--axial, vector--vector, and pseudoscalar--pseudoscalar coupling sectors, allowing the corresponding \(V_2^{AA}\), \(V_3^{AA}\), \(V_8^{AA}\), \(V_2^{VV}\), \(V_3^{VV}\), and \(V_3^{pp}\) interactions to be probed within a unified MRFM platform. Using representative parameters from existing MRFM experiments, we estimate the projected sensitivities under realistic noise conditions and compare them with current experimental limits. These results demonstrate that MRFM provides a promising platform for probing spin-dependent interactions at submicrometer length scales and has the potential to substantially improve existing constraints in the short-range regime.

\hspace{5mm}

\begin{acknowledgements}
\noindent{\it Acknowledgements}
This work was supported by the National Natural Science Foundation of China (Grants Nos. 12447128, 12475106 and T2388102), the China Postdoctoral Science Foundation (Grants No. 2024M760720), the National Key R\&D Program of China (Grant No. 2024YFF0727902), and the Fundamental Research Funds for the Central Universities (Grant No. JZ2025HGTB0176).
\end{acknowledgements}

\bibliography{main.bib}

\appendix
\counterwithout{equation}{section}
%%%%%%%%%% Merge with supplemental materials %%%%%%%%%%
\pagebreak
\widetext
\begin{center}
\textbf{\large Supplemental Materials: Searching for Exotic Spin-Dependent Interactions with Adiabatic Spin Inversion}
\end{center}
%%%%%%%%%% Merge with supplemental materials %%%%%%%%%%
%%%%%%%%%% Prefix a "S" to all equations, figures, tables and reset the counter %%%%%%%%%%
\setcounter{equation}{0}
\setcounter{figure}{0}
\setcounter{table}{0}
\makeatletter
\renewcommand{\theequation}{S\arabic{equation}}
\renewcommand{\thefigure}{S\arabic{figure}}
\renewcommand{\bibnumfmt}[1]{[#1]}
\renewcommand{\citenumfont}[1]{#1}
%%%%%%%%%% Prefix a "S" to all equations, figures, tables and reset the counter %%%%%%%%%%

\section{Spin-mechanical dynamics of MRFM}

\begin{figure}[htbp]
    \centering
    \includegraphics[width=0.45\textwidth]{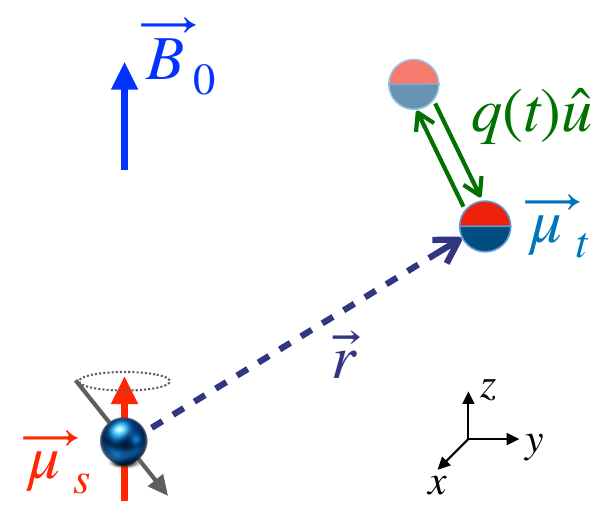}
    \caption{General geometry used in the analytical derivation. The ferromagnetic tip carries a magnetic moment \(\vec{\mu}_t\) aligned with the static magnetic field \(\vec B_0\) and undergoes relative motion with respect to the sample spin along an arbitrary direction \(\hat{u}\), with relative velocity \(\vec{v}=\dot q\,\hat{u}\). The electron spin $\vec{\mu}_s$ is polarized by \(\vec B_0\).}
    \label{fig:general}
\end{figure}

To keep the derivation independent of the specific MRFM geometry, we consider the general configuration illustrated in Fig.~\ref{fig:general}. The ferromagnetic tip carries a magnetic moment \(\vec{\mu}_t\) aligned with the static magnetic field \(\vec B_0\), and its relative motion with respect to the sample spin is taken to occur along an arbitrary direction \(\hat u\), with relative velocity \(\vec v=\dot q\,\hat u\), where \(q(t)\) denotes the relative displacement. The electron spin is polarized by \(\vec B_0\), such that its magnetic moment \(\vec\mu_s\) is aligned with \(\vec B_0\). The instantaneous relative position vector between the spin and the tip is \(\vec r\). The tip and the sample spin interact through the conventional magnetic dipole potential \(V_t\). If a hypothetical \(Z^\prime\)-mediated interaction exists, additional spin-dependent potentials \(V_2\), \(V_3\) and \(V_8\) arise between them. The cantilever-tip dynamics is described by the Lagrangian
\begin{align}
    \mathcal{L}(q,\dot{q})=\frac12 m_{\rm eff} \dot{q}^2-\frac12 m_{\rm eff}\omega_c^2 q^2 -V_t-V_2-V_3-V_8,
    \label{Lagq}
\end{align}
where $m_{\rm eff}$ is the effective motional mass of the cantilever-tip oscillator and $\omega_c$ is its resonance frequency. Using the relation \(\vec\sigma_{t,s}=-\vec\mu_{t,s}/\mu_B\), the interaction potentials can be recast in a Zeeman-like form,
\begin{align}
    V_i=-\vec{\mu}_s\cdot\vec{B}_i,\qquad i=t,2,3,8,
\end{align}
where \(\vec B_t\) is the magnetic field generated by the tip, while \(\vec B_2\), \(\vec B_3\), and \(\vec B_8\) denote the effective fields associated with the corresponding exotic interactions. They are given by
\begin{align}
    \vec{B}_t(r,\hat r)&=\frac{\mu_0}{4\pi r^3}\left[3(\vec{\mu}_t\cdot\hat{r})\hat{r}-\vec{\mu}_t\right], \\
    \vec{B}_2(r,\hat r)&=\frac{g_A^e g_A^e\hbar c}{4 \pi\mu_B^2}\vec{\mu}_t \frac{e^{-r / \lambda}}{r}, \\
    \vec{B}_3(r,\hat r)&=\frac{g_A^eg_A^e\hbar c}{4\pi\mu_B^2} \left[ A(r)\vec\mu_t-C(r)(\vec\mu_t\cdot\hat r)\hat r \right], \\
    \vec{B}_8(r,\hat r)&=\frac{g_A^e g_A^e\hbar}{8 \pi c\mu_B^2}(\vec{\mu}_t\cdot\vec{v})\vec{v}\frac{e^{-r / \lambda}}{r}.
    \label{Bfield}
\end{align}
Here the ferromagnetic tip is modeled as a continuous distribution of point magnetic dipoles. Since the tip dimensions in typical MRFM experiments are comparable to the tip–spin separation, the point magnetic dipole considered here represents an infinitesimal volume element of the tip rather than the entire tip.

To determine the MRFM response, two coupled aspects must be considered. The first is the generalized mechanical force generated by the interaction potential along the relative-motion coordinate. The second is the dynamics of the sample-spin polarization, which is modulated by the magnetic-field gradient and the exotic effective fields and, in turn, determines the time dependence of the force. We begin with the mechanical part. Applying the Euler--Lagrange equation to the Lagrangian in Eq.~\eqref{Lagq} gives
\begin{align}
    \ddot{q}+\frac{\omega_c}{Q}\dot{q}+\omega_c^2q=\frac{F_t+F_2+F_3+F_8}{m_{\rm eff}},
    \label{eomq}
\end{align}
where \(F_t\), \(F_2\), \(F_3\) and \(F_8\) denote the force components along the tip motion direction \(\hat{u}\), arising from the magnetic dipole interaction \(V_t\) and the exotic interactions \(V_2\), \(V_3\) and \(V_8\), respectively. Here, the term \(\omega_c\dot q/Q\) is phenomenologically introduced to account for the mechanical dissipation of the cantilever-tip oscillator, with \(Q\) denoting the mechanical quality factor. The corresponding force components are
\begin{align}
    F_t&=\frac{3\mu_0}{4\pi r^4}\Bigg[ (\vec{\mu}_t\cdot\vec{\mu}_s)(\hat{r}\cdot\hat{u}) - 3(\vec{\mu}_t\cdot\hat{r})(\vec{\mu}_s\cdot\hat{r})(\hat{r}\cdot\hat{u}) + (\vec{\mu}_t\cdot\vec{u}_{\perp})(\vec{\mu}_s\cdot\hat{r}) + (\vec{\mu}_t\cdot\hat{r})(\vec{\mu}_s\cdot\vec{u}_{\perp}) \Bigg], \label{gft} \\
    F_2&=\frac{g_A^eg_A^e \hbar c}{4\pi\mu_B^2} \vec{\mu}_t\cdot\vec{\mu}_s \frac{\D}{\D r}\left(\frac{e^{-r/\lambda}}{r}\right)\hat{r}\cdot\hat{u}, \label{gf2} \\
    F_3&=\frac{g_A^eg_A^e \hbar c}{4\pi\mu_B^2}\Bigg\{\frac{dA(r)}{dr}(\vec\mu_t\cdot\vec\mu_s)(\hat r\cdot\hat u)-\frac{dC(r)}{dr}(\vec\mu_t\cdot\hat r)(\vec\mu_s\cdot\hat r)(\hat r\cdot\hat u)-\frac{C(r)}{r}\Big[(\vec\mu_t\cdot\vec u_\perp)(\vec\mu_s\cdot\hat r)+(\vec\mu_t\cdot\hat r)(\vec\mu_s\cdot\vec u_\perp)\Big]\Bigg\}, \label{gf3} \\
    F_8&=-\frac{g_A^eg_A^e \hbar}{8\pi c \mu_B^2} \left\{ (\vec{\mu}_s\cdot\hat{u})(\vec{\mu}_t\cdot\hat{u})\left[ \dot{q}^2\frac{\D}{\D r}\left(\frac{e^{-r/\lambda}}{r}\right)\hat{r}\cdot\hat{u} + 2\ddot{q}\frac{e^{-r/\lambda}}{r}\right]+2(\dot{\vec{\mu}}_s\cdot\hat{u})(\vec{\mu}_t\cdot\hat{u})\dot{q}\frac{e^{-r/\lambda}}{r} \right\}, \label{gf8}
\end{align}
where $\vec{u}_{\perp}=\hat{u}-(\hat{u}\!\cdot\hat{r})\hat{r}$ is the component of $\hat{u}$ perpendicular to $\hat{r}$, and $\mu_B$ is the Bohr magneton.

We next determine how the sample-spin dynamics, driven by the magnetic-field gradient and the exotic effective fields, modulate the force acting on the tip. To evaluate this spin-mediated backaction in Eq.~\eqref{eomq}, the total field acting on the electron spin is written as
\begin{align}
    \vec{B}=\vec{B}_0+\vec{B}_t(r,\hat r)+\vec{B}_2(r,\hat r)+\vec{B}_3(r,\hat r)+\vec{B}_8(r,\hat r).
\end{align}
Since \(B_0\gg B_t,B_2,B_3,B_8\), the static field \(\vec B_0\) defines the spin quantization axis. To leading order, the magnetic and exotic fields enter the resonance condition through their components parallel to \(\vec B_0\). We therefore retain the longitudinal components \(B_t^z\), \(B_2^z\), \(B_3^z\) and \(B_8^z\) in the following analysis. Under the equilibrium longitudinal field \(B_0+\bar B_t^z+\bar B_2^z+\bar B_3^z\), the electron spin is split into two Zeeman levels. The microwave frequency $\omega_0$ is chosen to satisfy the resonance condition
\begin{equation}
    \omega_{0}=\gamma\left[B_{0}+\bar{B}^z_t+\bar{B}^z_2+\bar{B}^z_3\right],
    \label{rescond}
\end{equation}
where \(\gamma\) is the gyromagnetic ratio and the overbar denotes evaluation at the equilibrium position. The equilibrium contributions \(\bar B_2^z\) and \(\bar B_3^z\) are included because \(V_2\) and \(V_3\) generate nonzero static longitudinal effective fields. In contrast, \(V_8\) is velocity dependent and vanishes in the static limit, and therefore does not contribute to the equilibrium resonance condition. The spatial region satisfying Eq.~\eqref{rescond} forms the resonant slice. Since the exotic fields are treated perturbatively, the resonant-slice thickness is determined to leading order by the conventional tip-field gradient, $z_{\rm sl}\approx \delta\omega/(\gamma \vert\nabla B_t^z\vert)$ with $\delta\omega$ being the intrinsic resonance linewidth. As the cantilever oscillates, the resonant slice moves with the tip and periodically sweeps across the target spin. To describe the spin dynamics, we transform to a frame rotating about the direction of $\vec{B}_0$ at the frequency $\omega_0$. In this frame, the oscillating microwave field $B_1\cos(\omega_0t)$ appears as a static magnetic field perpendicular to $\vec{B}_0$, and the spin experiences the effective field $\vec{B}_{\rm eff}=B_1\vec{x}+\Delta B\vec{z}$, where $\Delta B = B_0 + B_{\rm t}^z + B_2^z + B^z_3 + B_8^z - \omega_0/\gamma$. Using Eq.~\eqref{rescond}, this becomes
\begin{align}
    \Delta B=\sum_{n=1}^{\infty}\frac{q^n}{n!}\left.\frac{\D^n \left(B^z_t+B_2^z+B_3^z\right)}{\D q^n}\right|_{q=0}+B^z_8,
    \label{detuB}
\end{align}
where the position-dependent static fields \(B_t^z\), \(B_2^z\), and \(B_3^z\) have been expanded about the equilibrium position. Neglecting spin relaxation over the timescale of interest, the spin dynamics in the rotating frame is governed by the Bloch equation
\begin{equation}
    \dot{\vec{\mu}}_s = \gamma \vec{\mu}_s \times \vec{B}_{\mathrm{eff}}.
    \label{preeom}
\end{equation}
Under the adiabatic condition \(\gamma B_{\rm eff}\gg\omega_c\), the spin adiabatically follows the effective field \(\vec B_{\rm eff}\), with its magnetic moment parallel to the field for the chosen adiabatic branch \cite{PhysRevA.66.023405}. In the small-detuning regime \(|\Delta B|\ll B_1\), the spin components can be approximated as
\begin{align}
    \mu_s^x &\approx \mu_s, \\
    \mu_s^y &= 0, \\
    \mu_s^z &= \mu_s^x \frac{\Delta B}{B_1}. \label{musz}
\end{align}
Within the additional small-displacement approximation, retaining only the linear (\(n=1\)) term in \(q\) from Eq.~\eqref{detuB} and substituting it into Eq.~\eqref{musz} yields
\begin{align}
    \mu_s^z&=\frac{\mu_s}{B_1}\Bigg\{\frac{-3\mu_0q}{4\pi r^4}\Big[3(\hat r\cdot\hat u)(\vec\mu_t\cdot\hat r)\hat r-(\hat r\cdot\hat u)\vec\mu_t-(\vec\mu_t\cdot\hat u_\perp)\hat r-(\vec\mu_t\cdot\hat r)\hat u_\perp\Big] \nonumber \\
    &+\frac{g_A^eg_A^e\hbar c}{4\pi\mu_B^2}q\,\frac{d}{dr}\left(\frac{e^{-r/\lambda}}{r}\right)(\hat r\cdot\hat u)\vec\mu_t \nonumber \\
    &+\frac{g_A^eg_A^e\hbar c}{4\pi\mu_B^2}q\Bigg[\frac{dA(r)}{dr}(\hat r\cdot\hat u)\vec\mu_t-\frac{dC(r)}{dr}(\vec\mu_t\cdot\hat r)(\hat r\cdot\hat u)\hat r-\frac{C(r)}{r}\left[(\vec\mu_t\cdot\hat u_\perp)\hat r+(\vec\mu_t\cdot\hat r)\hat u_\perp\right]\Bigg] \nonumber \\
    &+\frac{g_A^eg_A^e\hbar c}{4\pi\mu_B^2}\frac12\left(\frac{\dot q}{c}\right)^2(\vec\mu_t\cdot\hat u)\hat u\frac{e^{-r/\lambda}}{r}\Bigg\}\cdot\hat z.
    \label{muzexp}
\end{align}
Eq.~\eqref{muzexp} gives the linearized longitudinal spin polarization along the quantization axis defined by \(\vec B_0\), which governs the resonantly modulated MRFM signal and the spin-dependent backaction on the cantilever. Substituting this expression into Eq.~\eqref{eomq} yields the equation of motion for the cantilever displacement \(q(t)\).

The linear dependence of the longitudinal spin polarization on the cantilever displacement, \(\mu_s^z\propto q\), is valid only in the small-detuning regime \(|\Delta B|/B_1\ll1\). Beyond this regime, the longitudinal spin moment follows the full adiabatic response,
\begin{align}
    \mu_s^z=\mu_s\frac{\Delta B}{\sqrt{B_1^2+\Delta B^2}},
\end{align}
which satisfies the physical bound \(|\mu_s^z|\leq\mu_s\). To approximately account for this saturation, we consider harmonic cantilever motion \(q(t)=q_a\cos(\omega_ct)\), with \(\Delta B\simeq Gq\), where \(G=|\partial B_t^z/\partial q|\). By matching the fundamental harmonic of the nonlinear spin response to its linearized counterpart, we introduce an effective amplitude
\begin{align}
    q_{\rm eff}=q_aE(a),\qquad
    E(a)=\frac{4}{\pi}\int_0^{\pi/2}\frac{\cos^2\theta}{\sqrt{1+a^2\cos^2\theta}}\,d\theta,
    \quad a=\frac{Gq_a}{B_1},
    \label{qeff}
\end{align}
where \(\theta=\omega_ct\) denotes the oscillation phase. In the strongly modulated regime \(a\gg1\), \(q_{\rm eff}\) approaches \(4B_1/(\pi G)\), reflecting the saturation of the fundamental spin response. Here, \(q_{\rm eff}\) is introduced solely to correct the longitudinal spin polarization \(\mu_s^z\) in the linearized approximation, without modifying the actual cantilever motion. This treatment provides an order-of-magnitude approximation rather than a complete nonlinear calculation of the spin-mediated mechanical response.

\section{Signal of exotic interactions}

\begin{figure}
    \centering
    \includegraphics[width=0.45\linewidth]{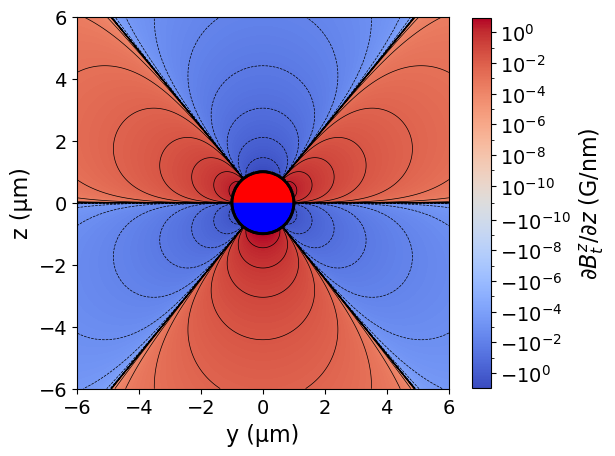}
    \includegraphics[width=0.45\linewidth]{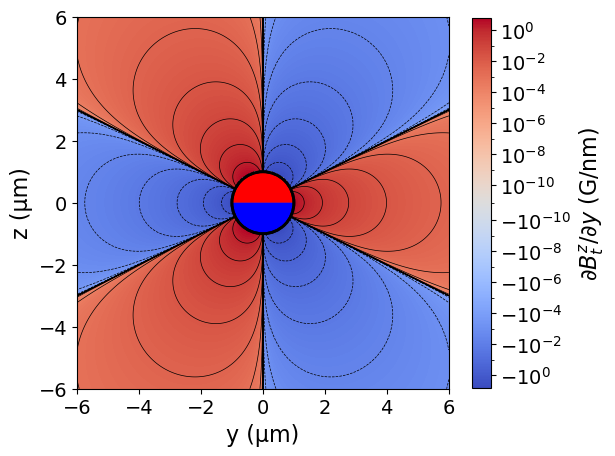}
    \caption{Spatial distributions of the magnetic-field gradients generated by a uniformly magnetized spherical ferromagnetic tip (\(r_t=1~\mu\mathrm{m}\), \(\mu_t=1.5\times10^{-12}~\mathrm{J/T}\)). Left: longitudinal gradient \(\partial B_t^z/\partial z\), relevant to the longitudinal geometry of Fig.~\ref{fig:setup}(b). Right: transverse gradient \(\partial B_t^z/\partial y\), relevant to the transverse geometry of Fig.~\ref{fig:setup}(c). The color maps and contours indicate the magnitude and sign of the corresponding gradients.}
    \label{fig:grad}
\end{figure}

The general formulation derived above applies to arbitrary cantilever motion. We now consider the two representative MRFM geometries shown in Fig.~\ref{fig:setup}. In the longitudinal configuration of Fig.~\ref{fig:setup}(b), the tip oscillates along the tip--sample separation direction, with the spin--mechanical coupling governed by the longitudinal gradient \(\partial B_t^z/\partial z\). In the transverse configuration of Fig.~\ref{fig:setup}(c), the tip oscillates perpendicular to this direction, and the coupling is determined by the transverse gradient \(\partial B_t^z/\partial y\). The spatial distributions of these two gradients generated by the ferromagnetic tip are shown in Fig.~\ref{fig:grad}, demonstrating that sizable magnetic-resonance-force coupling can be obtained in both configurations. To obtain transparent analytical expressions for the exotic-interaction signals, we focus on the small-detuning regime \(|\Delta B|/B_1\ll1\), where the longitudinal spin response can be linearized. We begin with the longitudinal configuration, in which both the tip--spin separation and the oscillation direction are aligned with \(\vec B_0\). The distance between the center of the tip magnet and the sample spin is denoted by \(z_0\). For clarity, we denote the displacement \(q\) by \(z\), emphasizing that the motion is along the direction of \(\vec{B}_0\). The finite geometry of the tip magnet needs to be taken into account. For simplicity while retaining the essential physics, we model the tip as a uniformly magnetized sphere with polarized electron density \(n_t\) and volume \(V_t\). Under the rigid-body approximation, all volume elements of the tip share the same velocity \(\dot z\) associated with the cantilever motion. Specializing Eqs.~\eqref{gft}, \eqref{gf2}, \eqref{gf3} and \eqref{gf8} to this geometry yields
\begin{align}
    F_t&=-\frac{3\mu_0}{2\pi}\frac{\mu_t\mu_s^z}{z_0^4}, \\
    F_2&=g_A^e g_A^e \frac{\hbar c}{4\pi \mu_B} \mu_s^z n_t \int_{V_t}\frac{\rm d}{\rm dr}\left( \frac{e^{-r/\lambda}}{r} \right)(\hat{r}\cdot\hat z){\rm d}V, \\
    F_3&=g_A^eg_A^e\frac{\hbar c}{4\pi\mu_B}\mu_s^z n_t\int_{V_t}\Bigg[\frac{dA(r)}{dr}(\hat r\cdot\hat z)-\frac{dC(r)}{dr}(\hat r\cdot\hat z)^3-\frac{2C(r)}{r}(\hat r\cdot\hat z)\left(1-(\hat r\cdot\hat z)^2\right)\Bigg]dV, \\
    F_8&=-g_A^e g_A^e \frac{\hbar}{4\pi c \mu_B} n_t \left[ \left( \mu_s^z \ddot{z}  + \dot{\mu}_s^z \dot{z} \right) \int_{V_t} \frac{e^{-r/\lambda}}{r} \D V + \frac12 \mu_s^z \dot{z}^2 \int_{V_t}\frac{\rm d}{\rm dr}\left( \frac{e^{-r/\lambda}}{r} \right)(\hat{r}\cdot\hat z) {\rm d}V \right]. \label{longF8}
\end{align}
For \(F_t\), the integration over the uniformly magnetized spherical tip can be evaluated analytically and is exactly equivalent to that of a point magnetic dipole at the sphere center. In contrast, the finite-range contributions \(F_2\), \(F_3\) and \(F_8\) are kept in their explicit volume-integral forms. For the present geometry, the longitudinal component of the sample-spin polarization along the quantization axis \(\vec B_0\), as given by Eq.~\eqref{muzexp}, takes the form
\begin{align}
    \mu_s^z=\frac{\mu_s}{B_1}\Bigg\{&-\frac{3\mu_0}{2\pi}\frac{\mu_t}{z_0^4}z+g_A^eg_A^e\frac{\hbar c}{4\pi\mu_B}n_t z\int_{V_t}\frac{\D}{\D r}\left(\frac{e^{-r/\lambda}}{r}\right)(\hat{r}\cdot\hat z)\D V \nonumber \\
    &+g_A^eg_A^e\frac{\hbar c}{4\pi\mu_B}n_t z\int_{V_t} \Bigg[\frac{dA(r)}{dr}(\hat r\cdot\hat z)-\frac{dC(r)}{dr}(\hat r\cdot\hat z)^3-\frac{2C(r)}{r}(\hat r\cdot\hat z)\left(1-(\hat r\cdot\hat z)^2\right)\Bigg] \D V \nonumber\\
    &+g_A^eg_A^e\frac{\hbar c}{8\pi\mu_B}n_t\left(\frac{\dot z}{c}\right)^2\int_{V_t}\frac{e^{-r/\lambda}}{r}\D V\Bigg\}.
    \label{longmuz}
\end{align}
Substituting \(\mu_s^z\) into the force terms \(F_i\), retaining only terms linear in \(g_A^eg_A^e\), and inserting the resulting forces into the tip equation of motion \eqref{eomq}, yields
\begin{align}
    \ddot{z} + (\omega_c^*)^2 z + \frac{\omega_c^*}{Q^*} \dot{z} &=\frac{F_2+F_3+F_8}{m_{\rm eff}},
\end{align}
with
\begin{align}
    F_2&=-g_A^e g_A^e \frac{3\hbar c \mu_0}{4\pi^2 \mu_B} \frac{\mu_t \mu_s n_t}{B_1} z_0^{-4} z \int_{V_t}\frac{\rm d}{\rm dr}\left( \frac{e^{-r/\lambda}}{r} \right)(\hat r\cdot\hat z) \D V, \\
    F_3&=-g_A^eg_A^e\frac{3\hbar c\mu_0}{4\pi^2 \mu_B} \frac{\mu_t\mu_sn_t}{B_1}z_0^{-4}z\int_{V_t} \Bigg[\frac{dA(r)}{dr}(\hat r\cdot\hat z)-\frac{dC(r)}{dr}(\hat r\cdot\hat z)^3-\frac{2C(r)}{r}(\hat r\cdot\hat z)\left(1-(\hat r\cdot\hat z)^2\right)\Bigg] \D V, \\
    F_8&=g_A^e g_A^e \frac{3\hbar\mu_0}{8\pi^2 c \mu_B} \frac{ \mu_t \mu_s n_t}{B_1} z_0^{-4} \left( z\ddot{z} + \frac12\dot{z}^2 \right) \int_{V_t} \frac{e^{-r/\lambda}}{r} \D V,
\end{align}
where the second term in the square brackets of Eq.~\eqref{longF8} is neglected because it is higher order in \(z/z_0\) and is therefore subleading in the small-displacement limit. Here, we have used the fact that the conventional magnetic dipole--dipole contribution in Eq.~\eqref{longmuz} gives \(\mu_s^z\propto z\), and hence \(F_t\propto z\). Therefore, this contribution is equivalent to a modification of the resonance frequency, namely \cite{PhysRevA.66.023405},
\begin{align}
    \omega_c^* &= \omega_c + \Delta\omega_c, \nonumber \\
    \Delta\omega_c &= -\frac{9\mu_0^2}{8\pi^2}\frac{\mu_t^2 \mu_s}{m_{\mathrm{eff}} \omega_c B_1} z_0^{-8}, \nonumber \\
    Q^* &= Q \left( 1 + \frac{\Delta\omega_c}{\omega_c} \right).
    \label{dipshif}
\end{align}
Since \(F_2,F_3\propto z\), the \(V_2\) and $V_3$ interaction produce a further shift of the cantilever resonance frequency, \(\omega_c^*\rightarrow\omega_c^*+\Delta\omega_2+\Delta\omega_3\), analogous to the shift induced by the conventional magnetic dipole interaction:
\begin{align}
    \Delta\omega_2 &= g_A^eg_A^e \frac{3 \hbar c \mu_0}{8\pi^2 \mu_B} \frac{ \mu_t \mu_s n_t}{ m_{\rm eff} \omega_c^* B_1} z_0^{-4}\int_{V_t} \frac{\D}{\D r}\left(\frac{e^{-r/\lambda}}{r}\right) (\hat r\cdot\hat z) \D V, \label{V2freshift} \\
    \Delta\omega_3 &= g_A^eg_A^e \frac{3 \hbar c \mu_0}{8\pi^2 \mu_B} \frac{ \mu_t \mu_s n_t}{ m_{\rm eff} \omega_c^* B_1} z_0^{-4}\int_{V_t} \Bigg[\frac{dA(r)}{dr}(\hat r\cdot\hat z)-\frac{dC(r)}{dr}(\hat r\cdot\hat z)^3-\frac{2C(r)}{r}(\hat r\cdot\hat z)\left(1-(\hat r\cdot\hat z)^2\right)\Bigg] \D V.
    \label{V3freshift}
\end{align}
The \(V_2,V_3\)-induced frequency shift and the conventional magnetic dipole contribution in Eq.~\eqref{dipshif} exhibit different dependences on the tip--sample separation \(z_0\). This difference provides a means to isolate the exotic contribution by measuring the cantilever resonance frequency at multiple tip--sample separations and fitting its distance dependence. In the absence of a statistically significant \(V_2,V_3\)-like component, the same analysis yields an upper bound on the coupling \(g_A^eg_A^e\). Experimentally, these small resonance-frequency shifts can be detected through phase-sensitive measurements of the cantilever displacement.

\begin{figure}[htbp]
    \centering
    \includegraphics[width=0.95\textwidth]{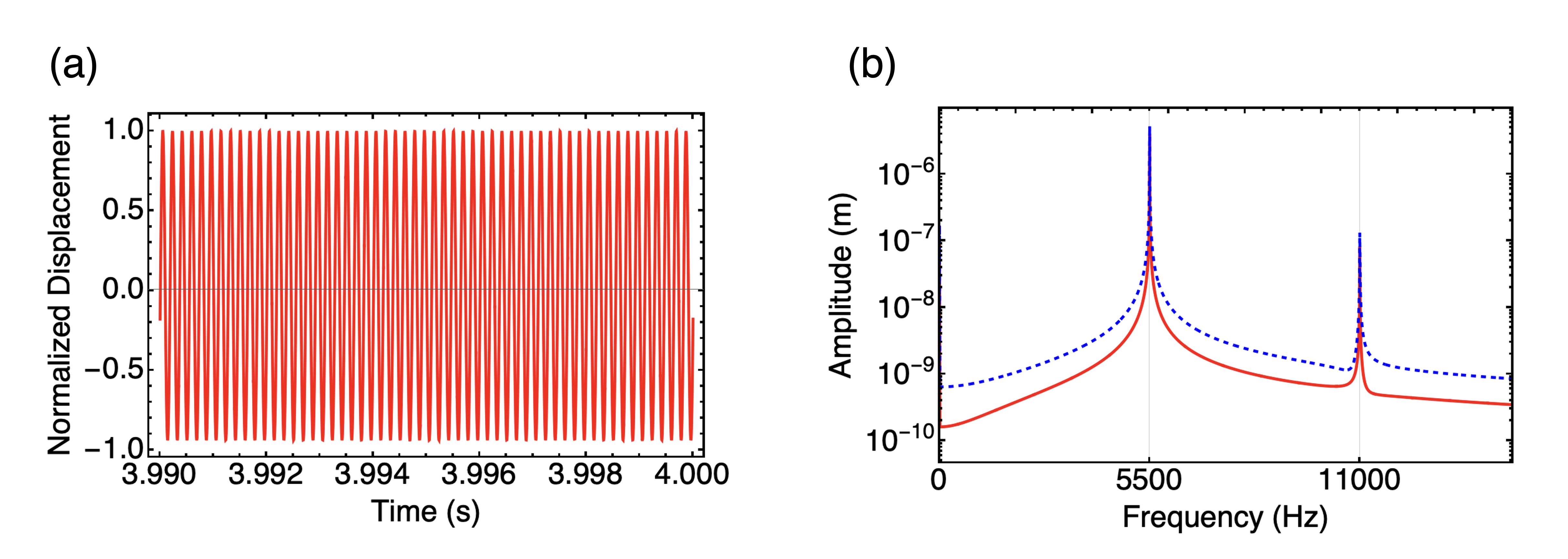}
    \caption{Comparison between numerical simulation and analytical approximation for the cantilever dynamics under $V_8$ interaction. (a) Numerical steady-state waveform from 3.99 s to 4.00 s, showing regular harmonic oscillation. (b) Combined FFT spectra: the red solid line represents the numerical simulation, while the blue dashed line represents the analytical approximation. Both spectra reveal a primary resonance peak at $\omega_{c} = 2\pi\times5500\ \mathrm{Hz}$ and a distinctive second-harmonic signature at $2\omega_{c}$ induced by the axial-vector interaction $V_8$. The high degree of overlap between the two curves validates the accuracy of our perturbative analytical model.}
    \label{fig:fft}
\end{figure}

For the \(V_8\) interaction, the situation is qualitatively different. Near resonance, the cantilever displacement follows \(z=z_a\cos(\omega_c t)\), yielding \(z\ddot z+(1/2)\dot z^2 \sim -(3/4)\omega_c^2 z_a^2\cos(2\omega_c t)\) after discarding the dc component. Consequently, the \(V_8\) interaction generates a force component at \(2\omega_c\), giving rise to a characteristic second-harmonic response. Since the conventional magnetic dipole-dipole interaction contributes predominantly at \(\omega_c\), this frequency separation provides a means of distinguishing the exotic signal from the dominant magnetic-resonance response. To illustrate the expected signal structure, rather than to simulate the complete noise-limited detection process, we numerically solve the cantilever equation of motion in the absence of stochastic noise. The parameters are chosen as \(\omega_c=2\pi\times5500~{\rm Hz}\) and \(Q=10^5\), while the remaining parameters are selected such that \(|F_8|/|F_t|=10^{-7}\). Starting from \(z=\dot z=0\), the cantilever is driven externally until steady-state oscillation is reached. Fig.~\ref{fig:fft}(a) shows the resulting time-domain displacement, while the solid curve in Fig.~\ref{fig:fft}(b) shows the corresponding displacement power spectrum obtained by Fourier transforming the time-domain signal. In addition to the dominant resonance peak at \(\omega_c\), a second peak appears at \(2\omega_c\), originating from the \(V_8\) interaction. The dashed curve represents the analytical approximation retaining only the second-harmonic term. Its agreement with the numerical solution illustrates the characteristic spectral signature of \(V_8\), while the projected detection sensitivity is evaluated separately using the noise model described in the sensitivity analysis.

Having analyzed the longitudinal geometry in Fig.~\ref{fig:setup}(b), we next specialize the general results to the transverse geometry shown in Fig.~\ref{fig:setup}(c), in which the relative oscillation velocity $q(t)\hat u=y(t)\hat y$ is perpendicular to \(\vec B_0\) and the tip is laterally displaced by $y_0$ from the sample spin. The distance vector between the center of tip magnet and the sample spin is \(\vec{r}_0=y_0\hat y+z_0\hat z\). A finite lateral offset \(y_0\neq0\) is required in this geometry, since it generates a nonzero gradient $\partial B_t^z/\partial y$ of the longitudinal tip field along the direction of cantilever motion. Specializing the general force expressions Eqs.~\eqref{gft}, \eqref{gf2}, \eqref{gf3} and \eqref{gf8} to this geometry gives
\begin{align}
    F_t&=\frac{3\mu_0}{4\pi}\frac{\mu_t\mu_s^z y_0(y_0^2-4z_0^2)}{r_0^7}, \\
    F_2&=g_A^e g_A^e\frac{\hbar c}{4\pi \mu_B} \mu_s^z n_t \int_{V_t}\frac{\rm d}{\rm dr}\left( \frac{e^{-r/\lambda}}{r} \right)(\hat r\cdot\hat y){\rm d}V, \\
    F_3&=g_A^eg_A^e\frac{\hbar c}{4\pi\mu_B}\mu_s^z n_t\int_{V_t}\left[\frac{\D A(r)}{\D r}-(\hat r\cdot\hat z)^2\frac{\D C(r)}{\D r}+\frac{2C(r)}{r}(\hat r\cdot\hat z)^2\right](\hat r\cdot\hat y)\D V, \\
    F_8&=0.
\end{align}
The vanishing of the \(V_8\) contribution follows directly from $\vec\mu_t\cdot\vec v=0$. For the present geometry, the longitudinal spin polarization obtained from Eq.~\eqref{muzexp} becomes
\begin{align}
    \mu_s^z&=\frac{\mu_s}{B_1}\Bigg\{\frac{3\mu_0}{4\pi}\frac{\mu_t y_0(y_0^2-4z_0^2)}{r_0^7}y+g_A^eg_A^e\frac{\hbar c}{4\pi\mu_B}n_t y \int_{V_t}\frac{\D}{\D r}\left(\frac{e^{-r/\lambda}}{r}\right)(\hat r\cdot\hat y)\D V \nonumber \\
    &+g_A^eg_A^e\frac{\hbar c}{4\pi\mu_B}n_t y\int_{V_t}\left[\frac{\D A(r)}{\D r}-(\hat r\cdot\hat z)^2\frac{\D C(r)}{\D r}+\frac{2C(r)}{r}(\hat r\cdot\hat z)^2\right](\hat r\cdot\hat y)\D V\Bigg\}.
\end{align}
Substituting \(\mu_s^z\) into the forces $F_t$ and $F_2$, retaining only terms linear in \(g_A^eg_A^e\), and inserting the resulting forces into the cantilever equation of motion \eqref{eomq} gives
\begin{align}
    \ddot{y} + (\omega_c^*)^2 y + \frac{\omega_c^*}{Q^*} \dot{y} &=\frac{F_2+F_3}{m_{\rm eff}},
\end{align}
with
\begin{align}
    F_2&=g_A^eg_A^e\frac{3\hbar c \mu_0}{8\pi^2 \mu_B}\frac{\mu_t\mu_sn_t}{B_1}\frac{y_0(y_0^2-4z_0^2)}{r_0^7}y\int_{V_t}\frac{\D}{\D r}\left(\frac{e^{-r/\lambda}}{r}\right)(\hat r\cdot\hat y)\D V, \nonumber \\
    F_3&=g_A^eg_A^e\frac{3\hbar c\mu_0}{8\pi^2\mu_B}\frac{\mu_t\mu_sn_t}{B_1}\frac{y_0(y_0^2-4z_0^2)}{r_0^7}y\int_{V_t}\left[\frac{\D A(r)}{\D r}-(\hat r\cdot\hat z)^2\frac{\D C(r)}{\D r}+\frac{2C(r)}{r}(\hat r\cdot\hat z)^2\right](\hat r\cdot\hat y)\D V.
\end{align}
Similar to the result obtained previously for the geometry in Fig.~\ref{fig:setup}(b), the conventional magnetic resonance force contribution in the first line is linear in the cantilever displacement \(y\) and can therefore be absorbed into a renormalized mechanical resonance frequency,
\begin{align}
    \omega_c^* &= \omega_c + \Delta\omega_c, \nonumber \\
    \Delta\omega_c &= - \frac{9\mu_0^2}{32\pi^2}  \frac{\mu_t^2\mu_s}{m_{\rm eff}\omega_cB_1} \frac{y_0^2(y_0^2-4z_0^2)^2}{r_0^{14}}, \nonumber \\
    Q^* &= Q \left( 1 + \frac{\Delta\omega_c}{\omega_c} \right).
    \label{dipshiftra}
\end{align}
As in the longitudinal configuration discussed above, the remaining \(V_2\) and \(V_3\) force is also proportional to the cantilever displacement, \(F_2,F_3\propto y\), and therefore produces an additional resonance-frequency shift \(\omega_c^*\rightarrow\omega_c^*+\Delta\omega_2+\Delta\omega_3\),
\begin{align}
    \Delta\omega_2&=-g_A^eg_A^e\frac{3\hbar c \mu_0}{16\pi^2 \mu_B}\frac{\mu_t\mu_s n_t}{m_{\rm eff}\omega_c^* B_1}\frac{y_0(y_0^2-4z_0^2)}{r_0^7}\int_{V_t}\frac{\D}{\D r}\left(\frac{e^{-r/\lambda}}{r}\right)(\hat r\cdot\hat y) \D V, \\
    \Delta\omega_3&=-g_A^eg_A^e\frac{3\hbar c \mu_0}{16\pi^2 \mu_B}\frac{\mu_t\mu_s n_t}{m_{\rm eff}\omega_c^* B_1}\frac{y_0(y_0^2-4z_0^2)}{r_0^7}\int_{V_t}\left[\frac{\D A(r)}{\D r}-(\hat r\cdot\hat z)^2\frac{\D C(r)}{\D r}+\frac{2C(r)}{r}(\hat r\cdot\hat z)^2\right](\hat r\cdot\hat y)\D V.
\end{align}
Thus, although the spin--mechanical coupling is generated by a transverse field gradient in the present geometry, both the conventional magnetic interaction and the \(V_2,V_3\) interactions lead to resonance-frequency shifts of the same form as those obtained for the longitudinal configuration in Fig.~1(b).

\section{Sensitivity estimation method}

For the $V_2$ and $V_3$ search, the sensitivity is ultimately limited by the uncertainty in measuring the cantilever resonance frequency. We estimate this uncertainty using a phase-sensitive detection scheme, as employed in the landmark MRFM experiments reported in Refs.~\cite{PhysRevLett.91.207604,PhysRevLett.87.277602,rugar2004single}. Near resonance, the cantilever motion can be written as $q(t)=q_a\cos(\omega_c t+\phi)$, where $q_a$ and $\phi$ denote the oscillation amplitude and phase, respectively. In this regime, displacement fluctuations are directly converted into phase fluctuations, with the corresponding power spectral densities (PSD) related by $S_{\phi}(\omega)=2S_q(\omega)/q_a^2$, where the factor of 2 accounts for the single-sided PSD convention. The phase response of the resonator is governed by its mechanical susceptibility according to Eq.~\eqref{eomq},
\begin{align}
    \chi(\omega)=\frac{1}{m_{\rm eff}\left( \omega_c^2-\omega^2-i\Gamma_c\omega \right)}, \ \Gamma_c\equiv\frac{\omega_c}{Q},
\end{align}
where $\Gamma_c$ denotes full width at half maximum (FWHM) of the resonance linewidth. The associated phase shift is $\phi=-\arctan\left[\Gamma_c\omega/(\omega_c^2-\omega^2)\right]$. Near the resonance frequency, the phase-frequency response becomes approximately linear,
\begin{align}
    \frac{\D\phi}{\D\omega}=-\frac{(\omega_c^2+\omega^2)\Gamma_c}{(\omega_c^2-\omega^2)^2+\Gamma_c^2\omega^2}.
\end{align}
We therefore have $\delta\phi\approx(2/\Gamma_c)\delta\omega$. Consequently, phase noise is converted into an equivalent resonance-frequency noise with PSD
\begin{align}
    S_\omega(\omega)=\frac{\Gamma_c^2}{4}S_\phi(\omega)=\frac{\Gamma_c^2}{2 q_a^2}S_q(\omega).
\end{align}
The uncertainty in the measured resonance frequency is determined by the integrated frequency-noise power, i.e.,
\begin{align}
    \sigma_{\omega_c}^2=\frac{1}{2\pi}\int_{-\pi f_B}^{\pi f_B} S_\omega(\omega)\D \omega=\frac{\Gamma_c^2\langle q^2\rangle}{2q_a^2},
\end{align}
where \(f_B\) corresponds to measurement bandwidth \(2\pi f_B\), and the displacement power \(\langle q^2\rangle\) is
\begin{align}
    \langle q^2\rangle=\frac{1}{2\pi}\int_{-\pi f_B}^{\pi f_B} S_q(\omega) \D\omega\approx\frac{S_F(\omega_c)}{4 m_{\rm eff}^2 \omega_c^2 \Gamma_c},
\end{align}
where \(S_F(\omega)\) and \(S_q(\omega)\) are the force-noise and displacement-noise PSDs, respectively, related by \(S_q(\omega)=|\chi(\omega)|^2S_F(\omega)\), with \(\chi(\omega)=\left[m_{\rm eff}\left(\omega_c^2-\omega^2+i\Gamma_c\omega\right)\right]^{-1}\) denoting the mechanical susceptibility of the cantilever. Here we assume that the integration bandwidth is comparable to the mechanical resonance linewidth, \(2\pi B\approx\Gamma_c\), and use the approximation \(S_F(\omega_c+\Gamma_c)\approx S_F(\omega_c)\). Since the displacement noise relevant to the frequency estimation is confined to the mechanical bandwidth $\Gamma_c$, for $T_{\rm int}\gg2\pi/\Gamma_c$ with $T_{\rm int}$ being the effective integration time, the measurement consists of $N=T_{\rm int}\Gamma_c/(2\pi)$ statistically independent realizations, reducing the variance by a factor of $N$. The resulting resonance-frequency uncertainty is therefore
\begin{align}
    \sigma_{\omega}\approx\sqrt{\pi}{\rm SNR}^{-1}\Gamma_c^{\frac12}T_{\rm int}^{-\frac12}, \ {\rm SNR}\equiv \frac{q_a}{\sqrt{\langle q^2\rangle}},
    \label{sigmafre}
\end{align}
where we have introduced the displacement signal-to-noise ratio ${\rm SNR}$. This result explicitly shows that the resonance-frequency resolution improves with increasing oscillation SNR and scales as $T_{\rm int}^{-1/2}$.

For the search of $V_8$, the sensitivity is determined by the uncertainty of the minimum detectable force. As discussed above, the $V_8$ interaction generates a force component at $2\omega_c$, allowing it to be separated spectrally from the magnetic resonance force background at $\omega_c$. The detection threshold is therefore set by the force noise of the cantilever. Let $S_F(\omega)$ denote the force-noise PSD, which is related to the displacement-noise PSD through the relation $S_q(\omega)=|\chi(\omega)|^2 S_F(\omega)$. Assuming that the magnetic-resonance force remains coherent over a characteristic time $\tau_m$, the corresponding measurement bandwidth is $\Delta\omega \approx 2\pi/\tau_m$. The force variance within this bandwidth is $\sqrt{2\pi S_F(\omega)/\tau_m}$. For a effective integration time $T_{\rm int}$, the number of statistically independent measurements is $N=T_{\rm int}/\tau_m$. Assuming white force noise, averaging reduces the variance of the measured force power by a factor of $N^{-1/2}$. Consequently, the uncertainty of the minimum detectable force scales as $N^{-1/4}$, yielding the expression at a given frequency $\omega$
\begin{align}
    \sigma_F(\omega)\approx \frac{\sqrt{2\pi S_F(\omega)}}{(T_{\rm int}\tau_m)^{1/4}}.
    \label{sigmaforce}
\end{align}
Unlike the \(V_2\) search, where the signal is extracted from phase-sensitive frequency measurements with an uncertainty scaling as \(N^{-1/2}\), the \(V_8\) signal is conservatively assumed to be detected through the excess power of a spectral peak at \(2\omega_c\), without phase-coherent averaging between successive measurement intervals. Consequently, the minimum detectable force amplitude scales as \(N^{-1/4}\), characteristic of incoherent power-spectral measurements. If the second-harmonic phase can be maintained or tracked, coherent demodulation could instead provide an improved amplitude sensitivity scaling as \(T_{\rm int}^{-1/2}\).

We next consider the signal extraction. For the $V_{2,3}$ interaction, Eqs.~\eqref{V2freshift}, \eqref{V3freshift} show that, at the scale $d\approx\lambda$ with $d$ being the characteristic source–sensor separation determined by the corresponding MRFM geometry, the induced resonance-frequency shift scales approximately as $\Delta\omega_c^{*}\propto d^{-6}$, whereas the frequency shift arising from the magnetic resonance force scales as $\Delta\omega_c\propto d^{-8}$. Likewise, the $V_8$ interaction produces a force oscillating at $2\omega_c$ with amplitude $F_8\propto d^{-5}$. The dominant background at the same frequency originates from the quadratic correction to the tip-induced residual longitudinal field $\Delta B$ in the rotating frame, proportional to $\D^2 B_t/\D q^2\vert_{q=0}q^2$ in Eq.~\eqref{detuB}, whose corresponding force scales as $d^{-9}$. These distinct distance dependence provide a direct means of separating the exotic interaction from the magnetic background. The corresponding distance-scaling analysis has been widely adopted in precision experiments to extract $Z^\prime$-mediated signals from dominant backgrounds \cite{Rong:2018yos,Jiao:2019ikv}.

To extract the exotic interaction signal, the resonance-frequency shift $\Delta\omega$ (for $V_{2,3}$) and the force excess $\Delta F$ in the $2\omega_c$ frequency bin of the force noise spectrum (for $V_8$) are measured at a series of tip-sample separations $d_i$ ($i=0,1,\ldots,N_d-1$), and are collectively denoted by $\Delta O(d_i)$. The measured data are then fitted to $\Delta O(d_i)=Cd_i^{-m}+Dd_i^{-n}$, where the first term represents the magnetic background, the second corresponds to the exotic interaction with $D\propto g_A^eg_A^e$, and the exponents $(m,n)$ take the values $(8,6)$ for the $V_{2,3}$ search and $(9,5)$ for the $V_8$ search. If no statistically significant exotic component is observed, the fitted uncertainty of $D$ is used to place an upper limit on the coupling $g_A^eg_A^e$. From the covariance matrix of the weighted least-squares fit, the standard error is $\sigma_D=\sqrt{\left[(X^{T}WX)^{-1}\right]_{DD}}$, where
\begin{align}
    X=\begin{pmatrix}
        d_0^{-m} & d_0^{-n}\\
        d_1^{-m} & d_1^{-n}\\
        \cdots & \cdots\\
        d_{N_d-1}^{-m} & d_{N_d-1}^{-n}
       \end{pmatrix},
    \qquad
    W=\frac{1}{\sigma_O^2} I_N.
\end{align}
Here $I_N$ is the $N_d\times N_d$ identity matrix, and $\sigma_O$ denotes the uncertainty of the measured observable at each distance point, namely $\sigma_\omega$ (Eq.~\eqref{sigmafre}) for $V_{2,3}$ and $\sigma_F(\omega)$ (Eq.~\eqref{sigmaforce}) for $V_8$. Explicitly, this gives
\begin{align}
    \sigma_D=\sigma_O\sqrt{\frac{\sum_i d_i^{-2m}}{\left(\sum_i d_i^{-2m}\right)\left(\sum_i d_i^{-2n}\right)-\left(\sum_i d_i^{-(m+n)}\right)^2}}.
\end{align}
For the expected null-signal case, we adopt a conservative two-sided 95\% confidence criterion for the signed fitted coefficient \(D\), giving \(|D|_{\rm lim}=1.96\sigma_D\). Using the relation $D=K g_A^e g_A^e$, where the proportionality constant $K$ is determined by the calculated $V_{2,3}$-induced frequency shift $\Delta\omega_{2,3}$ or the $V_8$-induced force $F_8$, the corresponding upper bound on the coupling can be obtained directly. Introducing the normalized distances $\rho_i=d_i/d_0$, the uncertainty of the fitted coefficient can be rewritten as
\begin{align}
    |K|g_A^e g_A^e d_0^{-n}\lesssim 1.96 \sigma_O R(\rho),\quad R(\rho)\equiv \sqrt{\frac{\sum_i \rho_i^{-2m}}
    {\left(\sum_i \rho_i^{-2m}\right)\left(\sum_i \rho_i^{-2n}\right)-\left(\sum_i \rho_i^{-(m+n)}\right)^2}}.
    \label{estiV2}
\end{align}
The geometric factor \(R(\rho)\) depends on both the number of measurements and their relative positions. Owing to the power-law structure characterized by the exponents \((m,n)\), the optimization of \(R(\rho)\) reduces to a two-distance design, in which the two tip--sample separations are chosen to maximize the distinguishability between the \(r_0^{-m}\) and \(r_0^{-n}\) dependences. Once these two separations are optimized, measurements at additional distinct distances do not introduce a new independent response direction and primarily improve the statistical precision. The optimal strategy is therefore to allocate repeated measurements, or equivalently the total integration time, between the two optimal separations. Taking the normalized separations to be \(\rho_0=1\) and \(\rho_1=d_1/d_0=\alpha\), and assigning fractions \(w\) and \(1-w\) of the total \(N_d\) measurements to them, respectively, the geometric factor becomes
\begin{align}
    R^2(w,\alpha)=\frac{w+(1-w)\alpha^{-2m}}{Nw(1-w)\left(\alpha^{-n}-\alpha^{-m}\right)^2}.
    \label{Fsq}
\end{align}
Minimization with respect to \(w\) yields
\begin{align}
    w_{\rm opt}=\frac{1}{1+\alpha^m}.
\end{align}
Substituting \(w_{\rm opt}\) into Eq.~\eqref{Fsq} and further minimizing with respect to \(\alpha\) gives the optimal distance ratio determined by
\begin{align}
    n\alpha_{\rm opt}^{m}-m\alpha_{\rm opt}^{n}=m-n.
\end{align}
For the \(V_{2,3}\) search, with \((m,n)=(8,6)\), this gives \(\alpha_{\rm opt}\simeq1.20\) and \(1.96R_{\min}\simeq23.6/\sqrt{N_d}\). For the \(V_8\) search, with \((m,n)=(9,5)\), one obtains \(\alpha_{\rm opt}\simeq1.21\) and \(1.96R_{\min}\simeq11.2/\sqrt{N_d}\). For the projected-sensitivity curves, we adopt a representative background-separation factor of \(1.96R_{\min}=10\) to illustrate the expected statistical penalty associated with simultaneously fitting the conventional magnetic background and the exotic signal. This factor is treated as an illustrative assumption rather than the result of a fully specified distance-sampling and integration-time allocation protocol.

The above sensitivity estimates assume idealized power-law distance dependences and thermal-noise-limited measurements. For a uniformly magnetized spherical tip, the external magnetic field is exactly dipolar, whereas deviations from spherical geometry and finite resonant-slice effects may modify the distance dependence. In particular, a relative distance error \(\delta d/d\) introduces a fractional error of approximately \(m\delta d/d\) in the magnetic background, while deviations from the assumed power-law exponent may produce a residual contribution to the fitted exotic-force coefficient. Accurate distance calibration and characterization of the magnetic background are therefore required. We further assume that fluctuations in the cantilever amplitude and resonance frequency, as well as other systematic contributions, remain below the statistical sensitivity. The projected constraints should thus be interpreted as thermal-noise-limited sensitivities under these assumptions.

\section{Noise budget}

We now analyze the dominant noise sources that determine the force sensitivity of the MRFM system. The total force-noise PSD acting on the cantilever can be written as
\begin{align}
  S_{FF}(\omega)=\left|\chi(\omega)\right|^{-2}S_{xx}^{\rm imp}+S_{FF}^{\rm ba}(\omega)+S_{FF}^{\rm th}(\omega)+S_{FF}^{\rm spin}(\omega),
\end{align}
where $S_{xx}^{\rm imp}$ denotes the displacement imprecision noise of the optical readout and $S_{FF}^{\rm ba}$ is the corresponding measurement backaction force noise. Together they constitute the readout noise and are constrained by the standard quantum limit through the uncertainty relation $S_{xx}^{\rm imp}S_{FF}^{\rm ba}\ge \hbar^2$, while $S_{FF}^{\rm th}$ and $S_{FF}^{\rm spin}$ represent the thermal force noise of the cantilever and the force noise arising from stochastic spin fluctuations in the sample, respectively, which together constitute the intrinsic force noise.

For a well-optimized resonant detector such as a state-of-the-art MRFM, the displacement sensitivity of the optical interferometer can be made sufficiently high that the readout noise remains below the intrinsic mechanical noise around the cantilever resonance. Consequently, the noise floor near $\omega_c$ is expected to be limited by intrinsic fluctuations rather than by the displacement readout. We further assume that a similar optimization can be achieved around $2\omega_c$, so that the projected sensitivities of both the $V_2$ and $V_8$ searches are ultimately limited by the intrinsic force noise. As a fundamental consequence of the fluctuation-dissipation theorem, the thermal force noise acting on the cantilever is white over the frequency range of interest \cite{PhysRevD.42.2437,RevModPhys.67.249}. Its force-noise PSD is given by
\begin{align}
    S_F^{\text{th}} = \frac{4 k_B T m_{\text{eff}} \omega_c}{Q},
    \label{eq:thermnoi}
\end{align}
where $T$ is the temperature. The second intrinsic noise source arises from statistical fluctuations of the spin polarization. For the $E^\prime$-center spin ensemble considered here, each defect can be treated as an independent electron spin with magnetic moment $\mu_B$. In thermal equilibrium under the magnetic field $B_0$, the populations of the two Zeeman levels follow the Boltzmann distribution,
\begin{align}
    p(\uparrow)=\frac{e^{\mu_B B_0 / k T}}{2 \cosh \left(\mu_B B_0 / k T\right)},\ p(\downarrow)=\frac{e^{-\mu_B B_0 / k T}}{2 \cosh \left(\mu_B B_0 / k T\right)}.
    \label{pol}
\end{align}
The corresponding expectation value and variance of the spin polarization are
\begin{align}
    \langle \mu_s \rangle&=N_e P\mu_B, \nonumber \\
    \left\langle \delta\mu_s^2\right\rangle&=\mu_B^2 N_e (1-P^2), \label{avevarM}
\end{align}
where $N_e$ is the effective number of participating electrons and the spin polarization rate $P$ is defined in Eq.~\eqref{pol} as
\begin{align}
    P\equiv \tanh(\mu_B B_0/kT).
\end{align}
When the rms fluctuation of the magnetization exceeds its thermal mean, \(\sqrt{\langle \delta\mu_s^2\rangle}>\langle \mu_s\rangle\), the spin polarization is dominated by statistical fluctuations rather than by the Boltzmann polarization. Using Eqs.~\eqref{avevarM}, this condition gives \(N_e<(1-P^2)/P^2\). For \(B_0=0.1~{\rm T}\) and \(T=2~{\rm K}\), the crossover occurs at \(N_e\sim10^3\), whereas our ensemble configuration has \(N_e\sim10^4\), for which the mean polarization exceeds its rms fluctuations. Nevertheless, spin fluctuations remain a source of noise, whose contribution at the measurement frequency is determined by their power spectral density, as discussed below. The magnetic resonance force itself originates from the mean polarized magnetization of the spin ensemble. For a total of \(N_e\) participating electrons, only a fraction \(P\) contributes to the net polarization, such that the effective magnetic moment entering the force expressions is \(\langle\mu_s\rangle\). Consequently, in the ensemble-type longitudinal geometry, the sample magnetic moment \(\mu_s\) represents the thermal expectation value \(\langle\mu_s\rangle\). The number of participating electrons \(N_e\) is determined by the density of \(E^\prime\)-center spins, \(n_e\), and the volume swept by the resonant slice during the cantilever oscillation. Approximating this volume as the product of the resonant-slice area \(A\) and the peak-to-peak cantilever displacement \(2z_a\), one obtains \(N_e\approx2n_eAz_a\).

Besides the fluctuation amplitude $\langle\delta\mu_s^2\rangle$, the corresponding noise spectrum also depends on the characteristic relaxation time of the spin ensemble. According to the fluctuation-dissipation theorem, both the variance and the correlation time are required to fully determine the noise PSD. In the MRFM context, the relevant timescale is the coherence time $\tau_m$ of the magnetic resonance force introduced above. Microscopically, $\tau_m$ characterizes the duration over which the spin reversals remain synchronized with the oscillation of the cantilever tip during a single measurement cycle. It is generally limited by spin relaxation and decoherence processes, including the rotating-frame spin-lattice relaxation time $T_{1\rho}$, the longitudinal relaxation time $T_1$, and the transverse dephasing time $T_2$. In the present work, rather than modeling these mechanisms explicitly, we adopt experimentally measured values of $\tau_m$ reported in MRFM experiments and use them directly in the sensitivity estimate. Assuming an exponentially decaying spin autocorrelation function,
\begin{align}
    \langle \delta\mu_s(0) \delta\mu_s(\tau) \rangle=
    \langle \delta\mu_s^2\rangle e^{-|\tau|/\tau_m},
\end{align}
the Wiener-Khinchin theorem yields the magnetic-moment noise PSD
\begin{align}
    S_{\mu}^{\rm spin}(\omega)=2\int \langle \delta\mu_s^2 \rangle e^{-|\tau|/\tau_m} e^{-i\omega\tau}\D \tau = \frac{4 \langle \delta\mu_s^2 \rangle \tau_m}{1+\omega^2\tau_m^2}.
\end{align}
The corresponding force noise PSD is
\begin{align}
    S_F^{\rm spin}(\omega)=G^2S_{\mu}^{\rm spin}(\omega)\approx\frac{9\mu_0^2 \mu_t^2 N_e \mu_B^2 \tau_m}{\pi^2 r_0^8 (1+\omega^2\tau_m^2)},
\end{align}
where we have used the approximation $\langle \delta\mu_s^2 \rangle\approx N_e\mu_B^2$ and the definition $G\equiv\partial B_t/\partial z|_{z=0}$ representing the magnetic field gradient of the tip magnet.Because the spin fluctuation noise PSD has a Lorentzian spectrum with a characteristic bandwidth of order \(1/\tau_m\), its contribution is strongly suppressed at signal frequencies satisfying \(\omega\tau_m\gg1\). For the baseline ensemble configuration, we obtain \(S_F^{\rm spin}/S_F^{\rm th}\approx3.3\times10^{-7}\) at \(\omega_c\) and \(8.2\times10^{-8}\) at \(2\omega_c\). The corresponding ratios for the optimized ensemble configuration are \(3.3\times10^{-4}\) and \(8.2\times10^{-5}\), respectively. Thus, within the adopted spin-fluctuation model, thermal force noise dominates over spin noise in both ensemble configurations, allowing us to approximate \(S_F(\omega)\approx S_F^{\rm th}\).

\section{Estimation of experimental parameters}

This Appendix summarizes the experimental parameters adopted in the sensitivity analysis and describes the estimation of the derived quantities listed in Table~\ref{tab:parameters}. To evaluate the proposed scheme under realistic conditions, we consider two representative MRFM experiments corresponding to single-spin detection in the transverse geometry of Fig.~\ref{fig:setup}(c) \cite{rugar2004single} and spin-ensemble detection in the longitudinal geometry of Fig.~\ref{fig:setup}(b) \cite{PhysRevLett.87.277602}.

The first parameter set is based on the single-spin MRFM experiment reported in Ref.~\cite{rugar2004single}. This experiment employs a small magnetic tip positioned close to a silica \(E^\prime\)-center sample. Combined with the low spin density of the sample, the resonant slice contains only a single electron spin, enabling single-spin detection. The effective mass is evaluated from the reported geometries and material properties of the cantilever and magnetic tip as \(m_{\rm eff}=0.24m_c+m_t\), where \(m_c\) and \(m_t\) are the masses of the cantilever and magnetic tip, respectively, and the coefficient 0.24 is the effective-mass factor of the fundamental flexural mode \cite{cleland2013foundations}. We also include the contribution of the long shaft used in the mass-loaded cantilever design to suppress higher-order mechanical modes. For analytical convenience, the magnetic tip is approximated by a uniformly magnetized sphere of comparable volume, with its magnetic moment \(\mu_t\) determined from the saturation magnetization of the tip material. The external magnetic field of this idealized sphere is equivalent to that of a point magnetic dipole at its center. For the spin-lattice relaxation time \(T_1\), we adopt the estimation method of Ref.~\cite{PhysRevLett.87.277602}, in which thermal currents in the electrically conductive magnetic tip constitute the dominant relaxation mechanism. All remaining parameters are taken from Ref.~\cite{rugar2004single} or, when not explicitly reported, from the companion experiment in Ref.~\cite{PhysRevLett.91.207604}.

The second parameter set is based on the spin-ensemble MRFM experiment of Ref.~\cite{PhysRevLett.87.277602}, in which a cantilever-mounted magnetic tip is coupled to an ensemble of silica \(E^\prime\) centers with a spin concentration of \(2\times10^{18}\,\mathrm{cm^{-3}}\). We adopt the experimentally measured spin synchronization time \(\tau_m\) at a tip--sample separation of \(z_0=1.5~\mu\mathrm{m}\). The effective mass is calculated using the same procedure described above. Although the original magnetic tip has a cylindrical geometry, we approximate it by a uniformly magnetized sphere of comparable volume while preserving its total magnetic moment \(\mu_t\). This simplification introduces geometry-dependent corrections but is adopted for the order-of-magnitude sensitivity estimates. The cross-sectional area of the resonant slice is estimated as \(A\approx0.25~\mu\mathrm{m}^2\), following the order of magnitude shown in Fig.~\ref{fig:grad}. We further assume that the polarized spins within this region contribute approximately equally to the spin--mechanical response, neglecting spatial variations in the magnetic-field gradient and exotic-interaction kernel. Such variations would require explicit integration over the resonant volume for a more accurate calculation.

For the spin-ensemble sensitivity estimates, we account approximately for the saturation of the longitudinal spin polarization by using the effective amplitude \(z_{\rm eff}\) defined in Eq.~\eqref{qeff}. Specifically, the displacement amplitude entering the linearized spin response in Eq.~\eqref{muzexp} is replaced by \(z_{\rm eff}\), and the resulting longitudinal spin polarization is substituted into the force expressions in Eqs.~\eqref{gft}--\eqref{gf8} to estimate the spin-dependent mechanical signals. This replacement applies only to the longitudinal spin response; the actual cantilever displacement, velocity, and acceleration remain determined by \(z_a\). In particular, the number of participating polarized spins is estimated from the swept resonant volume as \(N_s\simeq2n_eAz_aP\), where \(P\) is the thermal spin polarization. For the optimized ensemble configuration, we take \(T=30~\mathrm{mK}\) and \(z_a=100~\mathrm{nm}\), while keeping all other independent parameters unchanged. The effective amplitude \(z_{\rm eff}\) is recalculated to account for the saturated spin response, whereas the actual amplitude \(z_a\) is retained in estimating \(N_s\). This effective-response treatment provides an order-of-magnitude sensitivity estimate rather than a complete nonlinear calculation. Unless otherwise specified, we assume a total integration time of \(T_{\rm tot}=10^4~\mathrm{s}\) throughout the sensitivity analysis.

\end{document}